\documentclass[aps,twocolumn,pra]{revtex4-2}

\usepackage{amsfonts}
\usepackage{amssymb}
\usepackage{amsmath}
\usepackage{bm}
\usepackage{booktabs}
\usepackage{calc}
\usepackage{dcolumn}
\usepackage{graphicx}
\usepackage{xurl}
\usepackage[breaklinks=true,hidelinks]{hyperref}
\usepackage{comment}

\newcommand{\harr}{\leftrightarrow}

\usepackage{xcolor}

\def\be{ \begin{equation}}
\def\ee{ \end{equation}}
\def\bse{ \begin{subequations}}
\def\ese{ \end{subequations}}
\def\bea#1\ea{\begin{align}#1\end{align}}

\def\bi{\begin{itemize}}
\def\ei{\end{itemize}}

\def\bt{\begin{tabular}}
\def\et{\end{tabular}}

\def\ket#1{\vert #1 \rangle}
\def\bra#1{\langle #1 \vert}

\begin{document}

\title{Phase-Sensitive Benchmarking of Composite Quantum Gates with Chiral-Interference Circuits on Quantum Hardware}

\author{Georgi M. Aleksandrov}
\author{Nikolay V. Vitanov}
\affiliation{Center for Quantum Technologies, Department of Physics, Sofia University, 5 James Bourchier Boulevard, 1164 Sofia, Bulgaria}
\date{\today}

\begin{abstract}
We construct and experimentally implement compact gate-native circuits that simulate the state-transfer interference underlying three- and four-level chiral-resolution protocols. 
Both models are encoded in a two-qubit register, with the enantiomer-dependent sign of one of the couplings simulated by a conditional-phase operation in the four-level circuit and by the sign of a final rotation in the three-level circuit. 
On an IBM quantum processor, the two circuits produce the expected enantiomer-dependent output states with probabilities of nearly $98\%$. 
We then use these circuits as physically motivated, phase-sensitive benchmarks for composite quantum gates. 
We introduce rotation-angle error to the single-qubit operations and replace them by several composite gates, including B5, SK1, BB1, H5s, and X5. 
The comparison demonstrates that single-gate robustness does not translate to equivalent whole-circuit robustness. In particular, variable-rotation sequences do not preserve the required relative phases, making them unsuitable for error correction in circuits.
By contrast, the H5s/X5 sequences maintain high target-state populations for relative errors as large as \(50\%\), whereas elementary rotations reach the same threshold only for approximately \(8\%\). 
The three-level circuit exhibits a similar enhancement and additionally reveals an error-cancellation symmetry whose protection under composite replacement is exact only when the relevant full propagators satisfy an inverse relation. 
\end{abstract}

\maketitle

\section{Introduction}

Chirality is the property that prevents an object from being superimposed on its mirror image by rotations and translations alone. A chiral molecule, with its mirror image, are called \textit{enantiomers}. In an achiral environment, enantiomers have identical scalar properties. However, they interact differently with biological systems and other chiral surroundings. This makes  the separation and identification enantiomers particularly important in chemistry and pharmacology -- the two enantiomers of medications have different activity or side effects~\cite{penicillamine}.
Conventionally, separation techniques are primarilly chemical. They include crystallization, derivatization, kinetic resolution, and chiral chromatography. Despite their effectiveness, they are often slow, molecule specific and experimentally demanding~\cite{chiralreview,industry}.
The alternative physical approach relies on spectroscopy. Irradiating the molecules with optical or microwave pulses causes enantiomer-dependent population transfer at much faster time scales~\cite{BerovaNakanishi2000,Nafie2011,Barron2004,first_optical,shapiro,LiBruder2008}.

A widely used model for coherent chiral resolution is a closed loop of three states in which all three transitions are driven. 
The product of the three coupling amplitudes changes sign between the two enantiomers, so the phase accumulated around the loop differs by $\pi$. 
This phase difference changes the interference behavior between the excitation pathways from constructive to destructive, and can be converted into distinct final-state populations by a suitable pulse sequence~\cite{Hirota2012,YeZhangLi2018,LeibscherGiesenKoch2019}.
The technique of sensing chirality by exciting different rotational states using microwave fields is known as microwave three-wave mixing.~\cite{DomingosPerezSchnell2018}. 
Three-wave mixing was first demonstrated as reliable analysis of chiral mixtures~\cite{Patterson2013Nature,PattersonDoyle2013PRL}. The approach was expanded to broadband spectroscopy. Apart from enantiomers, it was also used to identify conformers~\cite{ShubertEtAl2014,ShubertEtAl2016}. 
Subsequently, the same process has been realized as a means to actively prepare specific states in a method called ``enantiomer-specific state transfer''~\cite{Eibenberger2017PRL}. Later studies laid the quantitative foundations~\cite{LeeEtAl2022PRL} and maintained separation in the presence of degenerate magnetic sublevels~\cite{LeibscherEtAl2022CommunPhys}. 
Further experiments demonstrated that the $\pi/2$-$\pi$-$\pi/2$ sequence is not required for achieving optimal transfer efficiencies~\cite{LeeEtAl2024NJP}. 
More recently, researchers have demonstrated transfer efficiencies above $92\%$ and state-specific enantiomeric purities of approximately $96\%$ in 1-indanol~\cite{experiment}.

Since chiral resolution relies on coherent interference, its performance is sensitive to pulse timing, detuning and relative field phases fluctuations. Different quantum-control methods have been employed to maintain robust interferometric contrast against those error sources. Adiabatic passage provides robust population transfer by following an instantaneous eigenstate~\cite{VitanovEtAl2017STIRAP}, while shortcuts to adiabaticity improve on adiabatic methods by adding an auxiliary coupling that counters the nonadiabatic transitions~\cite{ChenEtAl2010STA}. For chiral systems, one of the couplings has opposite sign. In that case, the ``counterdiabatic'' pulse acts counterdiabatically only for one enantiomer. This has been used to raise the chiral population contrast~\cite{VitanovDrewsen2019PRL}.

A simpler construction uses a sequential $\pi/2$-$\pi$-$\pi/2$ protocol, with one interaction shifted in phase by $\pi/2$~\cite{TorosovDrewsenVitanov2020PRA}. Because the transitions are addressed separately, the elementary rotations can be replaced by composite pulses designed to suppress systematic pulse-area or detuning errors. The same principle has been extended to single-Raman-single and Raman-single protocols using constant- and variable-rotation composite Raman sequences~\cite{TorosovDrewsenVitanov2020PRR}. Other recent approaches optimize the molecular fields more directly, including minimum-time control~\cite{StefanatosEtAl2025PRA}, robust microwave sequences for cyclic rotational manifolds~\cite{HongEtAl2026ChemPhysChem}, and Hamiltonian inverse engineering for four-level chiral systems~\cite{ZhangEtAl2026AQT}.

Composite pulse sequences have a long history for building robust quantum logic gates.
They have first been developed for NMR quantum computers~\cite{Cummins2003}, with applications to quantum counting~\cite{Xiao2006}. Since then, their use for high-fidelity gates has been demonstrated on trapped-ion~\cite{Mount2015} and superconducting quantum computers~\cite{Torosov2022}.

Their performance inside a circuit, however, cannot always be inferred from an isolated population-transfer measurement. 
An interferometric circuit depends on the complete action of each gate on a coherent superposition. 
Even though a composite pulse sequence may match the required population transfer with high precision, it can still fail within a larger circuit due to incorrect relative phases.
For that reason, we compare composite sequences through a phase-sensitive, physically relevant circuit.

Quantum computers allow for testing such protocols in a controlled environment by deliberately specifying the coherent gate errors.
Chiral discrimination has already been implemented in a superconducting transmon qutrit using nonadiabatic holonomic control~\cite{XuEtAl2024APL}. 
STIRAP and shortcut-to-adiabaticity schemes have also been translated to gate-based processors by discretizing the shaped fields and trotterizing the corresponding time-dependent Hamiltonian~\cite{chiral_stirap}. 
These approaches either reproduce the dynamics directly in a multilevel device or approximate the continuous-time molecular Hamiltonian digitally.

Here we take a complementary, gate-native approach. 
Instead of using Trotterization to simulate a complex time-dependent Hamiltonian evolution, we execute simple rotation gates representing the full action of single pulses, which do not overlap in time. Then sequences of pulses are chosen lead to chiral interference. The setup is mapped to compact two-qubit circuits. 

The first construction is more native to the quantum register. It maps four molecule levels to the four basis states of the two-qubit circuit. The enantiomer-dependent sign of one coupling is encoded by the existence or lack of a controlled-$Z$ gate. 

The second setup matches the established three-wave mixing. We use only three of the basis two-qubit states, leaving the fourth unoccupied according to the logical circuit. 
Particularly, we use an \(\mathrm{iSWAP}\) gate to transition between two of the molecular states. Here, the enantiomer-dependent sign is carried by the final single-qubit rotation. 
In both the three-level and four-level systems, the population starts from the same state for the S and R circuits but ends in a different final state, thereby simulating chiral discrimination.

The circuit construction constitutes the first contribution of this work. 
It provides compact gate-level representations of three- and four-level chiral interference and makes explicit the distinction between task-specific population equivalence and simulation of the complete molecular propagator. 
The second and central contribution is to use these chiral-interference circuits as physically motivated, phase-sensitive benchmarks for composite quantum gates. 
We introduce a correlated systematic error in the programmed single-qubit rotation angles and compare five-pulse variable-rotation sequences B5 with the constant-rotation sequences SK1, BB1, H5s, and X5. 
The comparison tests whether robustness obtained for an individual rotation survives inside a multigate protocol whose output depends on coherent phase accumulation and recombination. The central question is therefore not only whether a sequence suppresses an isolated transition-probability error, but whether its complete coherent propagator remains protected when the gate is embedded in an interferometric workload.

The paper is organized as follows. 
Section~\ref{sec:circuits} introduces the gate-native four- and three-level chiral-interference circuits and establishes their relation to the corresponding molecular state-transfer protocols. 
Section~\ref{sec:composite} defines the systematic-error model, the composite-gate constructions, and the experimental benchmarking procedure. 
Section~\ref{sec:results} presents the circuit-level robustness results and compares active composite-gate compensation with error cancellation arising from circuit symmetry. 
Section~\ref{sec:discussion} discusses the physical interpretation, scope, and limitations of the results, and Sec.~\ref{sec:conclusion} concludes the paper.
Appendices \ref{sec:a1} and \ref{sec:a2} present error analyses of the composite sequences used, respectively, as isolated sequences and in the circuits.

\section{Gate-Native Chiral-Interference Circuits \label{sec:circuits}}

Our objective is to construct the quantum circuits required by the selected chiral-discrimination protocols for a given common initial state for the S and R circuits and different final states. 
The two models are encoded in the computational basis $\{\ket{00},\ket{01},\ket{10},\ket{11}\}$ of a two-qubit register. 
Throughout this work, $\ket{xy}$ denotes the state in which the first qubit is in $\ket{x}$ and the second is in $\ket{y}$; this ordering is opposite to the bit-string convention used in Qiskit output. 
We use the single-qubit rotations
\bse
\begin{align}
R_k(\theta) &= e^{-i\theta\sigma_k/2},\quad k=x,y,z; \\
R(\theta,\phi) &= e^{-i\theta(\cos\phi\,\sigma_x+\sin\phi\,\sigma_y)/2},
\end{align}
\ese
and the relevant two-qubit gates.
The IBM Quantum processors offer the opportunity to test the single-qubit gates against rotation errors, which we will in the latter sections; unfortunately, no such option is available for the two-qubit gates.

\subsection{Four-level circuit}

\begin{figure*}[t]
    \centering
    \includegraphics[width=0.7\textwidth]{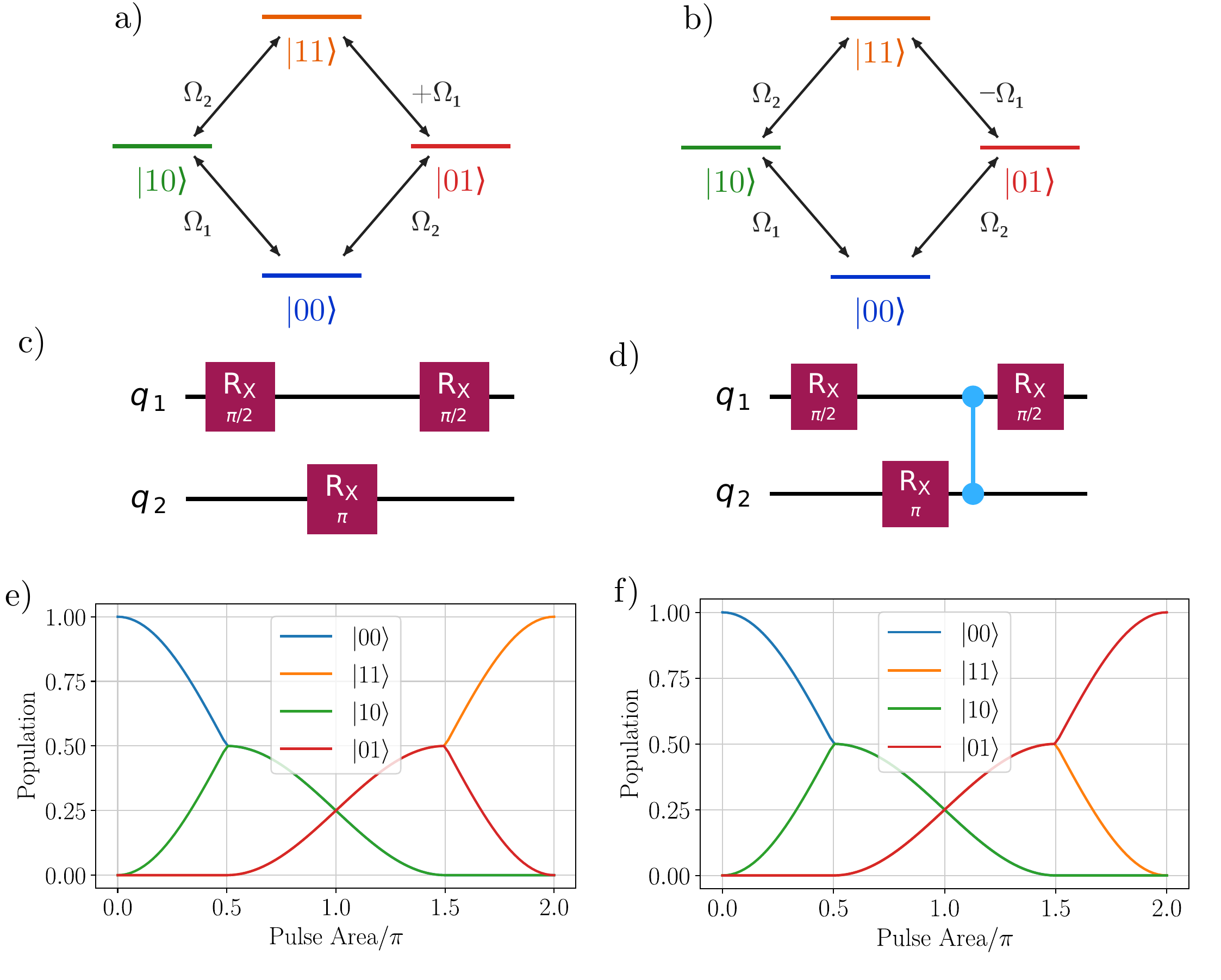}
    \caption{(a),(b) Interferometric scheme of the four-level model, with positive and negative signs of the loop coupling for the S- and R-circuits, respectively. (c),(d) Corresponding reduced quantum circuits. (e),(f) Ideal population evolution for the S- and R-circuits.}
    \label{fig:4ls_theory}
\end{figure*}

A four-level chiral interferometer maps directly onto the full Hilbert space of two qubits, see Fig.~\ref{fig:4ls_theory}. 
A rotation of the first qubit simultaneously drives the transitions $\ket{00}\harr\ket{10}$ and $\ket{01}\harr\ket{11}$ with the same effective Rabi frequency $\Omega_1$. 
A rotation of the second qubit similarly drives $\ket{00}\harr\ket{01}$ and $\ket{10}\harr\ket{11}$ with effective Rabi frequency $\Omega_2$. 
These linkages allow us to implement the interference protocol using single-qubit rotations together with a conditional phase gate that distinguishes the two model enantiomers.
However, we note that when designing the scheme, special care must be exercised for the steps in which the chiral resolution scheme requires driving a single transition; this is achieved by making sure that the paired parallel transition is unpopulated during this gate.

We consider a model in which the coupling on the $\ket{01}\harr\ket{11}$ transition changes sign for the R-circuit, so that its effective Rabi frequency is $-\Omega_1$ rather than $\Omega_1$ [Fig.~\ref{fig:4ls_theory}(b)]. 
For a pulse on the first qubit, the operation with equal coupling signs on the two transitions is
\be
U_{S}^{(1)}(\theta)=R_x(\theta)\otimes I.
\label{eq:four-us}
\ee
The corresponding operation with the sign reversed only in the $\ket{01}\harr\ket{11}$ subspace is
\be
U_{R}^{(1)}(\theta)=R_x(\theta)\otimes\ket{0}\!\bra{0}+R_x(-\theta)\otimes\ket{1}\!\bra{1}.
\label{eq:four-ur}
\ee
This conditional signed rotation is obtained exactly by conjugating the single-qubit rotation with CZ gates,
\be
U_{R}^{(1)}(\theta)=\mathrm{CZ}\bigl[R_x(\theta)\otimes I\bigr]\mathrm{CZ}.
\label{eq:four-cz-conjugation}
\ee

A complete implementation on an arbitrary two-qubit input would use the conjugated form in Eq.~\eqref{eq:four-cz-conjugation} for every pulse whose upper transition changes sign. 
The present state-transfer task admits a shorter circuit. 
The system starts in $\ket{00}$, so the sign of the unoccupied $\ket{01}\harr\ket{11}$ transition is irrelevant during the first $\pi/2$ pulse. 
For the final signed rotation, the second CZ in the exact conjugation acts only after all coherent operations and changes no computational-basis probability. 
It can therefore be omitted for the population measurement considered here.

Starting from $\ket{00}$, the protocol consists of an $R_x(\pi/2)$ rotation on qubit 1, an $R_x(\pi)$ rotation on qubit 2, a CZ gate for the R-circuit only, and a final $R_x(\pi/2)$ rotation on qubit 1 [Fig.~\ref{fig:4ls_theory}(c),(d)]. The CZ gate leaves the first three computational states unchanged and transforms $\ket{11}$ into $-\ket{11}$.

Let $U_R^{\mathrm{th}}$ denote the complete theoretical sequence constructed from the exact conditional signed rotations, and let $U_R^{\mathrm{exp}}$ denote the reduced one-CZ circuit used experimentally. 
These operators are not identical on an arbitrary input,
$U_R^{\mathrm{th}}\neq U_R^{\mathrm{exp}}$.
For the prescribed input, however, they give identical computational-basis probabilities,
\be
\begin{aligned}
\left|\bra{j}U_R^{\mathrm{th}}\ket{00}\right|^2
&=\left|\bra{j}U_R^{\mathrm{exp}}\ket{00}\right|^2,\\
&\hspace{-1.8em}j\in\{00,01,10,11\}.
\end{aligned}
\label{eq:four-population-equivalence}
\ee
The reduced circuit is therefore population equivalent to the complete four-level protocol for the state-transfer task considered here. 
The only reason to omit the second CZ gate is to avoid the error introduced by this gate into the circuit.
The resulting interference is constructive in different output channels for the two circuits and guides them to a different final state. 
As summarized in Table~\ref{tab:four-level-evolution}, the S circuit ends in $-\ket{11}$, whereas the R circuit ends in $-i\ket{01}$.

\begin{table}[tb]
\caption{Ideal state evolution in the four-level circuit. 
The CZ operation is applied only in the circuit representing the R-enantiomer. 
}
\label{tab:four-level-evolution}
\centering
\begin{tabular}{|l|l|l|}
\hline
\textbf{Operation} & \textbf{State (S)} & \textbf{State (R)} \\
\hline
Initialize & $\ket{00}$ & $\ket{00}$ \\
$R_x(\pi/2)\otimes I$ & $\frac{1}{\sqrt{2}}(\ket{00}-i\ket{10})$ & $\frac{1}{\sqrt{2}}(\ket{00}-i\ket{10})$ \\
$I\otimes R_x(\pi)$ & $\frac{1}{\sqrt{2}}(-i\ket{01}-\ket{11})$ & $\frac{1}{\sqrt{2}}(-i\ket{01}-\ket{11})$ \\
CZ (R only) & $\frac{1}{\sqrt{2}}(-i\ket{01}-\ket{11})$ & $\frac{1}{\sqrt{2}}(-i\ket{01}+\ket{11})$ \\
$R_x(\pi/2)\otimes I$ & $-\ket{11}$ & $-i\ket{01}$ \\
\hline
\end{tabular}
\end{table}

\subsection{Three-level circuit}

\begin{figure*}[tbph]
    \centering
    \includegraphics[width=0.7\textwidth]{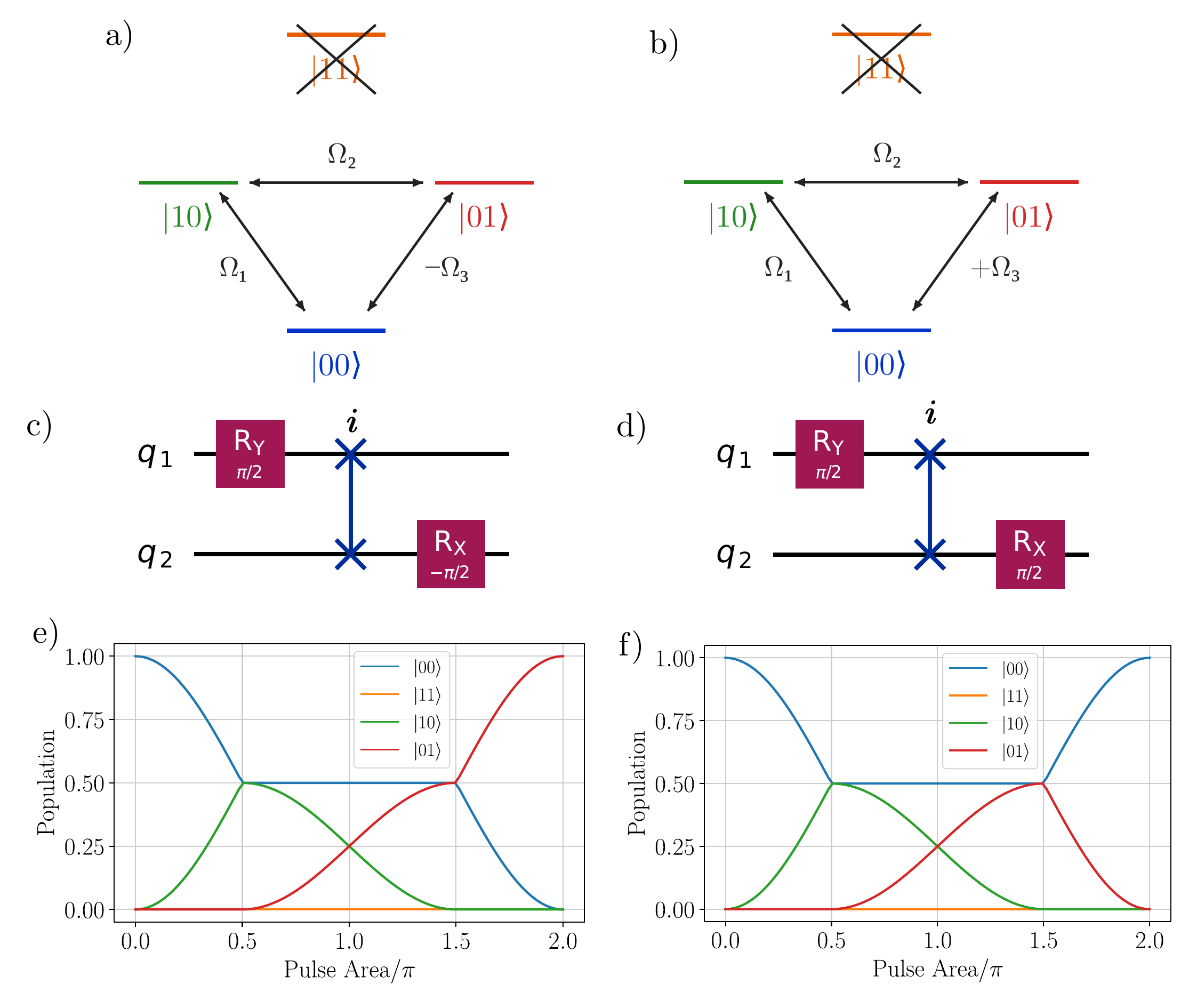}
    \caption{(a),(b) Interferometric scheme of the three-level model, with positive and negative signs of the loop coupling for the S- and R-circuits, respectively. (c),(d) Corresponding \(\mathrm{iSWAP}\)-based quantum circuits. (e),(f) Ideal population evolution for the S- and R-circuits.}
    \label{fig:3ls_theory}
\end{figure*}

A closed three-state loop provides the most direct physical setting for enantioselective population transfer. 
For a chiral molecule, the electric-dipole moment has three components, $\vec\mu=(\mu_a,\mu_b,\mu_c)$, and the pseudoscalar product $\mu_a\mu_b\mu_c$ changes sign under spatial reflection. 
Consequently, when the three transitions are coupled through the corresponding dipole components, the two enantiomers acquire opposite signs around the interaction loop. 
The protocol considered here uses sequential rather than simultaneous pulses and is equivalent to the scheme already implemented experimentally~\cite{experiment}.

The gate-based representation is less direct than in the four-level model because a two-qubit register contains four computational states, whereas the molecular model contains only three; hence we have to take special care to restrict the dynamics to three states only. 
We encode the effective levels in $\ket{00}$, $\ket{01}$, and $\ket{10}$, while requiring the auxiliary state $\ket{11}$ to remain unpopulated at the logical gate boundaries of the ideal evolution. 
We choose the mapping such that the $\ket{00}\harr\ket{01}$ coupling changes sign for the R-circuit.

Starting from \(\ket{00}\), the molecular sequence consists of a \(\pi/2\) pulse on \(\ket{00}\harr\ket{10}\), a \(\pi\) pulse on \(\ket{10}\harr\ket{01}\), and a \(\pi/2\) pulse on \(\ket{00}\harr\ket{01}\). Ideally, the S branch is transferred completely to \(\ket{01}\), whereas the R branch returns to \(\ket{00}\). The implemented circuit uses an \(R_y(\pi/2)\) rotation on qubit 1, an \(\mathrm{iSWAP}\) gate, and an \(R_x(\pm\pi/2)\) rotation on qubit 2, where the negative sign represents the S branch [Fig.~\ref{fig:3ls_theory}(c),(d)]. With the rotation convention defined above, \(\mathrm{iSWAP}\) acts as \(R_x(-\pi)\), rather than \(R_x(\pi)\), in the single-excitation subspace \(\{\ket{01},\ket{10}\}\). The corresponding phase difference relative to a nominal \(+\pi\) transition pulse is compensated by the phase of the final rotation for the prescribed state-transfer task. 
The corresponding state evolution is summarized in Table~\ref{tab:three-level-evolution}.

\begin{table}[tb]
\caption{Ideal state evolution in the \(\mathrm{iSWAP}\)-based three-level circuit. 
The sign of the final rotation is negative for the S-circuit and positive for the R-circuit. For the input $\ket{00}$, the sequence is population equivalent to one containing a transition-selective resonant $\pi$ pulse, although the corresponding unitaries differ by phases in the $\{\ket{01},\ket{10}\}$ subspace.}
\label{tab:three-level-evolution}
\centering
\begin{tabular}{|l|l|l|}
\hline
\textbf{Operation} & \textbf{State (S)} & \textbf{State (R)}\\
\hline
Initialize & $\ket{00}$ & $\ket{00}$\\
$R_y(\pi/2)\otimes I$ & $\frac{1}{\sqrt{2}}(\ket{00}+\ket{10})$ & $\frac{1}{\sqrt{2}}(\ket{00}+\ket{10})$ \\
iSWAP & $\frac{1}{\sqrt{2}}(\ket{00}+i \ket{01})$ & $\frac{1}{\sqrt{2}}(\ket{00}+i \ket{01})$ \\
$I\otimes R_x(\pm\pi/2)$ & $i \ket{01}$ & $\ket{00}$ \\
\hline
\end{tabular}
\end{table}

The first $R_y(\pi/2)$ rotation formally acts on both $\ket{00}\harr\ket{10}$ and $\ket{01}\harr\ket{11}$. 
At that stage, however, neither $\ket{01}$ nor $\ket{11}$ is populated, so only the intended transition contributes. 
After the \(\mathrm{iSWAP}\) gate, the state lies entirely in the span of \(\ket{00}\) and \(\ket{01}\). 
The final rotation therefore acts only on $\ket{00}\harr\ket{01}$ because the amplitudes of $\ket{10}$ and $\ket{11}$ vanish. 
Thus, at every logical gate boundary of the ideal circuit, the auxiliary state $\ket{11}$ has zero amplitude. 

With the definition \(R_x(\theta)=\exp(-i\theta\sigma_x/2)\), the implemented \(\mathrm{iSWAP}\) exchange is
\be
U_{\mathrm{iSWAP}}=\ket{00}\!\bra{00}+\ket{11}\!\bra{11}+i\ket{01}\!\bra{10}+i\ket{10}\!\bra{01}.
\label{eq:selective-pi}
\ee
Because \(R_x(\pi)=-i\sigma_x\), Eq.~\eqref{eq:selective-pi} corresponds to \(R_x(-\pi)\) in the \(\{\ket{01},\ket{10}\}\) subspace. An exact unitary realization of a transition-selective \(+\pi\) rotation with the convention defined above would instead use \(\mathrm{iSWAP}^{\dagger}\). The implemented \(\mathrm{iSWAP}\) circuit is nevertheless population equivalent to the desired three-level state-transfer protocol for the prescribed input \(\ket{00}\), after the corresponding final-pulse phase is chosen.

\subsection{Baseline hardware realization}

The two circuit families were executed on IBM's \texttt{ibm\_kingston} processor, based on the Heron r2 architecture. 
Each point was estimated from $2000$ shots without measurement-error mitigation. For the four-level circuits, the ideal output channels were observed with populations of $98.4\%$ in $\ket{11}$ for the S-circuit and $98\%$ in $\ket{01}$ for the R-circuit. The shot noise at that level is $0.3\%$. 
For the three-level circuits, the ideal-case populations are $98.4\%$ in $\ket{01}$ for the S-circuit and $97.7\%$ in $\ket{00}$ for the R-circuit. The average population across all measurements from non-ideal gates in the R-circuit is $98.3\%$. Since the average aggregates $10^5$ total measurements, its ``shot noise'' error is only $0.04\%$. 

Having established their ideal operation, we will present the results from their experimental testing in Sec.~\ref{sec:results} and will use them to test whether the robustness of composite rotations survives at the level of the complete circuit.

\section{Composite-Gate Benchmarking Protocol\label{sec:composite}}

We emulate a correlated systematic error in the single-qubit rotation angles by applying the replacement
\be
R(\theta,\phi)\longrightarrow R\bigl[(1+\varepsilon)\theta,\phi\bigr],
\label{eq:error-model}
\ee
where $\varepsilon$ is the common fractional over- or under-rotation. In a molecular implementation, the same unitary model describes a systematic error in the Rabi frequency or the pulse duration.

In the four-level circuit, Eq.~\eqref{eq:error-model} is applied to the two nominal \(\pi/2\) rotations and to the nominal \(\pi\) rotation. In the three-level circuit, it is applied only to the two nominal \(\pi/2\) rotations; the middle \(\mathrm{iSWAP}\) operation is kept fixed. The CZ and \(\mathrm{iSWAP}\) gates are not protected by the composite constructions considered here. The benchmark therefore isolates one correlated coherent error channel in the single-qubit operations rather than attempting general processor-level error mitigation. In the present implementation, \(\varepsilon\) is introduced through the programmed logical rotation angles before transpilation; it should therefore be interpreted as a controlled digital emulation of a correlated coherent over- or under-rotation, rather than as a deliberate analog miscalibration of the calibrated hardware pulses.

The elementary rotations are replaced by the composite sequences listed in Table~\ref{tab:composites}. The H5s sequence implements the nominal $\pi/2$ rotations, while X5 implements the nominal $\pi$ rotation~\cite{PhysRevA.104.012609}. We compare this pair with variable-rotation B5 sequences~\cite{torosov2011,torosov2019}, and the constant-rotation sequences SK1~\cite{wimperis1994}, and BB1~\cite{BrownHarrowChuang2004}.

\begin{table}[t]
\centering
\caption{Areas and phases of the individual pulses in the composite sequences used in the benchmark (see also Appendix \ref{sec:a1}).}
\label{tab:composites}
\begin{tabular}{|c|c|c|}
\hline
\textbf{Sequence} & \textbf{Areas/$\pi$} & \textbf{Phases/$\pi$} \\
\hline
H5s ($\pi/2$) & $0.45,1,1,1,0.45$ & $1.9494, 0.5106, 1.3179, 0.5106,$ \\
 & & 1.9494 \\
X5 ($\pi$) & $1,1,1,1,1$ & $0.0672, 0.3854, 1.1364, 0.3854$, \\
 & & 0.0672 \\
\hline
B5 ($\pi/2$) & $0.5,1,1,1,0.5$ & $0,3/8,3/2,11/8,0$ \\
B5 ($\pi$) & $1,1,1,1,1$ & $0,0.8,0.4,0.8,0$ \\
\hline
SK1 ($\pi/2$) & $0.5,2,2$ & $0,0.539893,-0.539893$ \\
SK1 ($\pi$) & $1,2,2$ & $0,0.580431,-0.580431$ \\
\hline
BB1 ($\pi/2$) & $0.5,1,2,1$ & $0,0.539893,1.619679,0.539893$ \\
BB1 ($\pi$) & $1,1,2,1$ & $0,0.580431,1.741292,0.580431$ \\
\hline
\end{tabular}
\end{table}

The chiral-interference circuits provide a stricter test than isolated population transfer. Their output depends on the relative phase of the intermediate superposition as well as on the populations produced by the constituent gates. A variable-rotation sequence may reproduce the desired transition probability without reproducing the complete target unitary. Such a sequence can therefore appear robust in an isolated population measurement and still distort the final interference signal. By contrast, constant or universal composite rotations are designed to preserve the coherent transformation required by the surrounding circuit.

Table~\ref{tab:specifications} summarizes the device parameters and circuit resources. Both experiments use transpiler optimization level 0, and $2000$ shots per circuit. The qubits were selected for having relatively low SPAM and gate errors at the time before submitting the job. No measurement-error mitigation is applied. The error sweeps consist of \(510\) circuits per job. The composite implementations substantially increase the logical and native gate counts, which provides a direct test of whether coherent-error suppression outweighs the additional circuit overhead.

\begin{table}[t]
\caption{Technical specifications of the two hardware experiments. The labels \(q_1\) and \(q_2\) refer, respectively, to the first and second qubits listed in the ``Qubits used'' row; paired readout values follow the same order.}
\label{tab:specifications}
\centering
\begin{tabular}{|l|c|c|}
\hline
\textbf{Data} & \begin{tabular}{@{}c@{}}\textbf{Four-level}\\ \textbf{circuit}\end{tabular} & \begin{tabular}{@{}c@{}}\textbf{Three-level}\\ \textbf{circuit}\end{tabular}\\
\hline
Qiskit version & \multicolumn{2}{c}{2.5.0}\\ \hline
Execution date & Jul 14, 2026 & Jul 17, 2026 \\ \hline
Transpiler opt. level & \multicolumn{2}{c}{0} \\ \hline
Qubits used & 88, 89 & 148, 149 \\ \hline
CZ error & $1.572\times10^{-3}$ & $1.977\times10^{-3}$ \\ \hline
\(\sqrt{X}\) error (\(q_1\)) & $3.102\times10^{-4}$ & $3.579\times10^{-4}$ \\ \hline
\(\sqrt{X}\) error (\(q_2\)) & $1.857\times10^{-4}$ & $1.801\times10^{-4}$ \\ \hline
\(T_1(q_1)\) & $226.96~\mathrm{\mu s}$ & $236.16~\mathrm{\mu s}$\\ \hline
\(T_1(q_2)\) & $169.52~\mathrm{\mu s}$ & $179.37~\mathrm{\mu s}$ \\ \hline
\(T_2(q_1)\) & $19.76~\mathrm{\mu s}$ & $219.44~\mathrm{\mu s}$ \\ \hline
\(T_2(q_2)\) & $123.9~\mathrm{\mu s}$ & $205.7~\mathrm{\mu s}$ \\ \hline
Readout \(P(1|0)\) & \bt{c} 0.684\%\\0.586\%\et & \bt{c}0.439\%\\0.464\%\et \\ \hline
Readout \(P(0|1)\) & \bt{c} 0.122\%\\0.024\%\et & \bt{c}0.195\%\\0.024\%\et \\ \hline
\multicolumn{3}{|c|}{\textbf{Gate counts and execution}}\\ \hline
Logical (elementary) & S: 3, R: 4 & 3\\ \hline
Logical (H5s) & S: 15, R: 16 & 11 \\ \hline
Logical (Variable rotation) & S: 15, R: 16 & 11 \\ \hline
Logical (SK1) & S: 9, R: 10 & 7 \\ \hline
Logical (BB1) & S: 12, R: 13 & 9 \\ \hline
Native (elementary) & S: 15, R: 16 & 32 \\ \hline
Native (H5s) & S: 75, R: 76 & 72 \\ \hline
Native (Variable rotation) & S: 75, R: 76 & 72 \\ \hline
Native (SK1) & S: 45, R: 46 & 52 \\ \hline
Native (BB1) & S: 60, R: 61 & 62 \\ \hline
Circuit depth (elementary) & S: 10, R: 11 & 22\\ \hline
Circuit depth (H5s) & S: 50, R: 51 & 62 \\ \hline
Circuit depth (Variable rot.) & S: 50, R: 51 & 62 \\ \hline
Circuit depth (SK1) & S: 30, R: 31 & 42 \\ \hline
Circuit depth (BB1) & S: 40, R: 41 & 52 \\ \hline
Job duration & $270~\mathrm{s}$ & $270~\mathrm{s}$\\ \hline
Circuits per job & 510 & 510 \\ \hline
Shots per circuit & 2000 & 2000 \\ \hline
\begin{tabular}{@{}c@{}}Measurement-error\\ mitigation\end{tabular} & \multicolumn{2}{c|}{None} \\
\hline
\end{tabular}
\end{table}

The principal benchmark observable is the population of the ideal target state: $\ket{11}$ and $\ket{01}$ for the S and R circuits of the four-level circuit, and $\ket{01}$ and $\ket{00}$ for the S and R circuits of the three-level system. 
We use a target-state population of \(94\%\) as a common reporting threshold for comparing the width of the robust region. This level lies several percentage points below the approximately \(98\%\) zero-error hardware ceiling and is used only to quote a common robustness width; it does not enter the theoretical calculations. 
Experimental uncertainty bands are proportional to twice the shot-noise standard deviation,
\be
2\sigma_p=2\sqrt{\frac{p(1-p)}{N_{\mathrm{shots}}}},
\label{eq:shot-noise}
\ee
where $p$ is the corresponding population.

The complete single-gate propagator expansions are derived in Appendix~\ref{app:single-gate-errors}, and their propagation through the three- and four-level circuits is worked out in Appendix~\ref{app:circuit-errors}. 
The perturbative expressions identify the local error orders near $\varepsilon=0$, whereas the solid curves in
Figs.~\ref{fig:compare_composite}-\ref{fig:3ls_experiment} are obtained by exact multiplication of the erroneous gate propagators over the full error interval. 
To compare the error-dependent shapes independently of the fixed hardware offset, each ideal theoretical curve is multiplied by the corresponding measured target-state population at \(\varepsilon=0\), for which the ideal prediction is unity. This introduces only a single vertical normalization and no error-dependent fit parameter; the predicted dependence on \(\varepsilon\) is unchanged.

\section{Circuit-Level Benchmark Results\label{sec:results}}

\subsection{Four-level circuit}

\begin{figure*}[t]
    \centering
    \includegraphics[width=\textwidth]{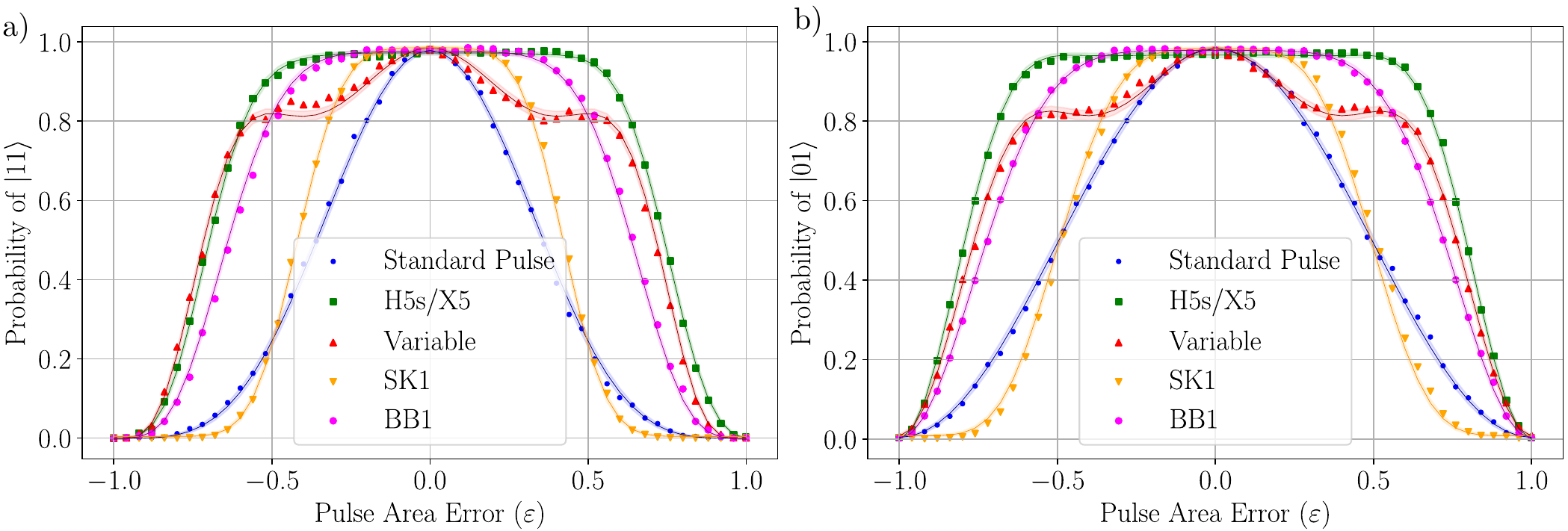}
    \caption{Target-state populations of the (a) S and (b) R circuits of the four-level circuit as functions of the correlated rotation-angle error $\varepsilon$. Elementary rotations are compared with H5s/X5, B5 variable rotations, SK1, and BB1. The points are hardware data from \texttt{ibm\_kingston}; the solid lines are scaled theoretical predictions and the shaded bands represent twice the shot-noise uncertainty.}
    \label{fig:compare_composite}
\end{figure*}

The results from the experimental implementation of the four-level circuits are presented in Fig.~\ref{fig:compare_composite}.
The elementary implementation is strongly affected by a common over- or under-rotation. 
Its target-state population remains above $94\%$ for both circuits only for approximately $|\varepsilon|\lesssim 0.08$. 
Replacing the elementary gates by composite rotations substantially broadens the response, for all composite sequences.
The H5s/X5 implementation maintains target-state populations above $94\%$ for errors in the interval $\varepsilon\in[-0.44,0.52]$ in both enantiomeric circuits, see Fig.~\ref{fig:compare_composite}.

The increasing flatness of the central regions in Fig.~\ref{fig:compare_composite} follows directly from the circuit-level expansions in Appendix~\ref{app:circuit-errors}. 
Assuming perfect two-qubit gates, the target-population error is quadratic in $\varepsilon$ for the elementary and B5 implementations, quartic for SK1, and sixth order for BB1 and H5s/X5. 
This hierarchy reflects compensation of the complete single-qubit propagator to progressively higher order.

The comparison also demonstrates why the robustness of isolated population-transfer is not sufficient for circuit operation. 
The B5 variable-rotation sequences reproduce the target transition probabilities but do not stabilize the complete gate phases. 
They therefore provide the weakest circuit-level improvement among the composite constructions. SK1 and BB1 preserve the coherent transformation more effectively, while the H5s/X5 pair gives the broadest high-population region in the present experiment.

More explicitly, the isolated variable-rotation B5 $\pi/2$ and $\pi$ transition probabilities have errors beginning at orders $\varepsilon^8$ and $\varepsilon^{10}$, respectively, but the complete B5 $\pi/2$ propagator retains a first-order coherent error. 
Recombination of the two interferometric pathways converts this residual phase error into a quadratic target-population error for the complete circuit. 
The B5 response is therefore considerably narrower than that of the constant-rotation composite gates despite its high-order isolated population compensation.

The two enantiomeric branches need not have identical residual error profiles, because the conditional phase changes how the single-gate errors interfere at the output. 
For the H5s/X5 implementation, the H5s contribution cancels exactly from the target amplitude of the four-level R circuit, so that $P_{4R}^{\mathrm{H5s/X5}}=P_{\mathrm{X5},\pi}$. 
The R-branch theoretical curve is therefore even in $\varepsilon$. 
Both H5s and X5 contribute in the S branch, where higher-order odd terms generate a weak asymmetry between over- and under-rotation errors. 
BB1 similarly has a sixth-order leading error but also acquires a seventh-order circuit contribution, so its two sides need not be exactly symmetric.

\subsection{Three-level circuit and conditional symmetry protection}

\begin{figure*}[t]
    \centering
    \includegraphics[width=\textwidth]{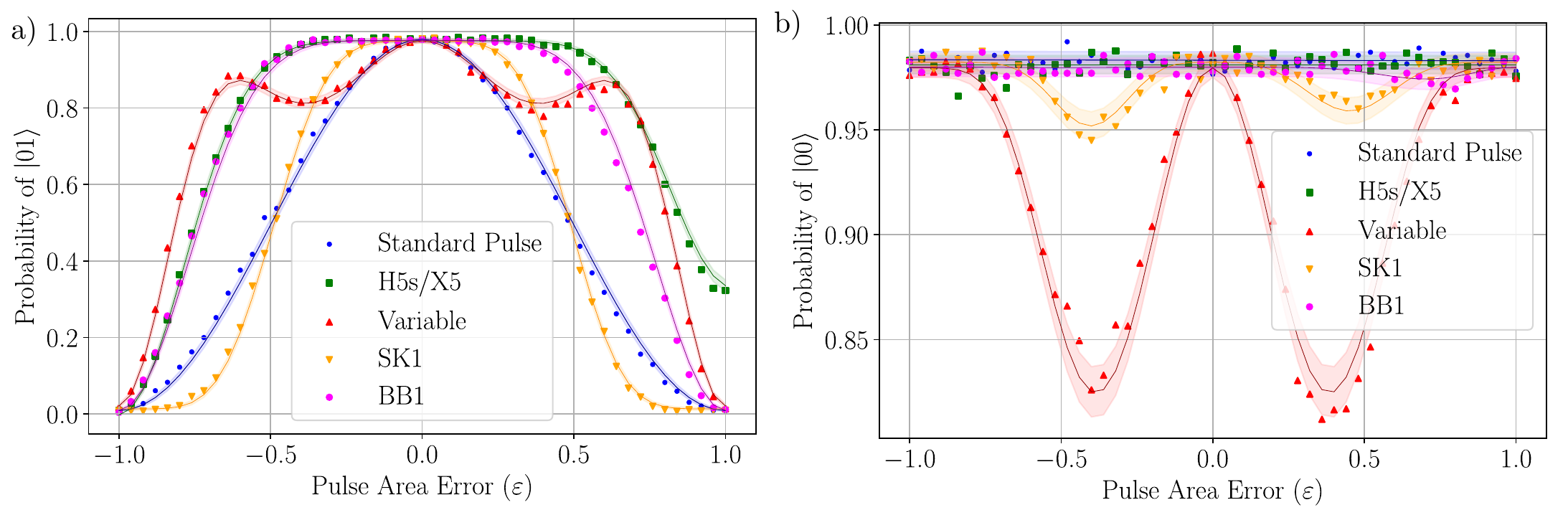}
    \caption{Target-state populations of the (a) S and (b) R branches of the three-level circuit as functions of the correlated rotation-angle error $\varepsilon$. The target states are $\ket{01}$ and $\ket{00}$, respectively. 
    Elementary pulses are compared with H5s, variable rotation, SK1 and BB1 composite $\pi/2$ rotations. For the R enantiomer, elementary pulses exhibit exact cancellation of the common rotation-angle error. For composite replacements, this protection persists only when the final erroneous composite propagator is the exact inverse of the transported first propagator; asymmetric sequences can break this condition.}
    \label{fig:3ls_experiment}
\end{figure*}

Because the middle operation is a fixed \(\mathrm{iSWAP}\), the imposed error acts only on the two nominal \(\pi/2\) rotations.
In the S circuit, replacing these rotations by H5s substantially broadens the high-population region. 
The measured population in \(\ket{01}\) remains above \(94\%\) in the interval \(\varepsilon\in[-0.44,0.52]\) with H5s, whereas the elementary circuit reaches the same threshold only up to approximately \(|\varepsilon|\lesssim0.08\), similarly to the four-level scenario.

The R circuit provides a qualitatively different test. For elementary rotations, the \(\mathrm{iSWAP}\) transports the first erroneous \(R_y\) rotation to an effective \(R_x\) rotation with the opposite nominal angle, so the common rotation-angle error cancels against the final rotation. For a composite replacement, however, this symmetry protection is conditional: exact cancellation requires the complete final erroneous composite propagator to be the inverse of the transported first composite propagator. Merely programming the opposite nominal target angle is not sufficient for a generic asymmetric sequence, because inverting a product also requires reversing the pulse order. This explains why the asymmetric B5 sequence can show pronounced deviations in Fig.~\ref{fig:3ls_experiment}(b), whereas the symmetric H5s construction preserves the cancellation.

For the nearly flat elementary and symmetry-preserving H5s data sets, the mean \(\ket{00}\) populations over the sweep are both approximately \(98.3\%\). Their difference is not resolved within the precision of the present experiment; this comparison should therefore be interpreted as a test of the overhead added by the longer symmetry-preserving implementation, not as evidence that every composite realization is symmetry protected.

Within the assumed systematic-error model, the cancellation is exact to all orders only under the exact-inverse condition stated above. Appendix~\ref{app:circuit-errors} derives this condition explicitly for the implemented \(\mathrm{iSWAP}\) circuit and shows how asymmetric composite sequences violate it.

\subsection{Noise contributions and resource overhead}

The four-level noise-model decomposition is summarized in Table~\ref{tab:noise-breakdown-four}. The IBM-derived noise model gives slightly larger deviations than observed on the QPU---\(1.9\%\) for the elementary circuit and \(2.4\)--\(3.4\%\) for the composite circuits---and is used here to separate the main modeled contributions. For the elementary circuit, the dominant modeled contribution is SPAM/readout error, whereas for the longest five-pulse composite implementations the thermal-relaxation/decoherence contribution becomes comparable to or larger than the SPAM contribution. In this data set one selected qubit has a \(T_2\) approximately one seventh of the device median (Table~\ref{tab:specifications}); it was nevertheless selected because of its low readout and gate errors.

\begin{table}[t]
\caption{Deviation from the ideal four-level result for different components of the IBM noise model.}
\label{tab:noise-breakdown-four}
\centering
\begin{tabular}{|l|c|c|c|c|c|}
\hline
\textbf{Component} & \multicolumn{5}{c|}{\textbf{Deviation}}\\
\hline
 & Elementary & X5/H5s & Variable & SK1 & BB1\\
\hline
SPAM & $1.3\%$ & $1.4\%$ & $1.4\%$ & $1.5\%$ & $1.3\%$ \\
Gate errors & $0.2\%$ & $0.8\%$ & $0.9\%$ & $0.5\%$ & $0.7\%$ \\
Thermal relax. & $0.5\%$ & $1.7\%$ & $1.7\%$ & $1.0\%$ & $1.2\%$ \\
\hline
Full noise profile & $1.9\%$ & $3.3\%$ & $3.4\%$ & $2.4\%$ & $3.1\%$ \\
\hline
QPU observation & $1.8\%$ & $2.8\%$ & $2.5\%$ & $2.0\%$ & $2.0\%$ \\
\hline
\end{tabular}
\end{table}

For the three-level R circuit, the full noise-model simulation gives generally consistent errors with the hardware result, within the margin of the shot noise. The decomposition in Table~\ref{tab:noise-breakdown-three} identifies SPAM as the dominant contribution, followed by gate errors. The iSWAP is transpiled into two native CZ gates, each with error of about $0.2\%$ (see Table~\ref{tab:specifications}), 14 RZ gates and 6 $\sqrt X$ gates.

\begin{table}[t]
\caption{Deviation from the ideal three-level result for different components of the IBM noise model.}
\label{tab:noise-breakdown-three}
\centering
\begin{tabular}{|l|c|c|c|c|c|}
\hline
\textbf{Component} & \multicolumn{5}{c|}{\textbf{Deviation}}\\
\hline
 & Elementary & X5/H5s & Variable & SK1 & BB1\\
\hline
SPAM & $1\%$ & $1.1\%$ & $1.3\%$ & $1.1\%$ & $1.0\%$ \\
Gate errors & $0.7\%$ & $1.1\%$ & $0.7\%$ & $0.9\%$ & $1.1\%$ \\
Thermal relax. & $0.2\%$ & $0.2\%$ & $0.3\%$ & $0.1\%$ & $0.2\%$ \\
\hline
Full noise profile & $1.6\%$ & $2.2\%$ & $1.7\%$ & $1.9\%$ & $2.2\%$ \\
\hline
QPU observation & $2.0\%$ & $1.9\%$ & $1.6\%$ & $1.9\%$ & $2.1\%$ \\
\hline
\end{tabular}
\end{table}

The resource increase is substantial: the longest four-level composite circuits contain approximately five times more native single-qubit operations and reach total transpiled depths of \(50\)--\(51\) instead of \(10\)--\(11\), while the corresponding three-level total depth increases from \(22\) to \(62\) (Table~\ref{tab:specifications}). The added decoherence penalty is therefore not negligible in an absolute sense, as also indicated by Tables~\ref{tab:noise-breakdown-four} and \ref{tab:noise-breakdown-three}; nevertheless, over most of the tested coherent-error interval it remains smaller than the improvement obtained by suppressing the imposed systematic rotation error. This is consistent with the processor coherence times and the reported single-qubit gate duration of approximately \(32~\mathrm{ns}\)~\cite{ibm_kingston}.

\section{Discussion\label{sec:discussion}}

The first contribution of this work is the construction of compact two-qubit circuits that implement the interference mechanism underlying chiral discrimination. 
In the four-level model, the full two-qubit Hilbert space is used and the enantiomer-dependent sign is represented by a controlled phase gate. 
In the three-level model, the molecular loop is encoded in \(\ket{00}\), \(\ket{01}\), and \(\ket{10}\), with an \(\mathrm{iSWAP}\) exchanging the latter two states and the handedness-dependent sign carried by the final rotation. 
In both cases, reversing the sign around the interaction loop redirects constructive interference into a different output channel.

Direct compilation keeps this physical structure visible. Instead of discretizing a shaped pulse and trotterizing a time-dependent Hamiltonian, each stage of the state-transfer protocol is represented by a small set of gates. 
The outcome of this compactness is that the circuits are tailored to the input state $\ket{00}$ and to different final states in the computational basis. 
Omitting the final conditional phase in the four-level circuit and using the implemented \(\mathrm{iSWAP}\), with the corresponding final-pulse phase, in place of the exact transition-selective \(+\pi\) exchange in the three-level circuit leave the measured populations unchanged for this state-transfer task, but do not preserve the action on arbitrary input states.
A complete unitary implementation would require the full conditional rotations in the four-level model and a phase-preserving $\mathrm{iSWAP}^{\dagger}$-type exchange, together with the corresponding final-pulse phase.
However, this distinction is immaterial for chiral resolution, where the sole objective is to prepare both enantiomers in the same initial state and drive them to distinct final states.

The central experimental result is that the advantage of full-rotation composite gates survives inside a complete interferometric circuit. The circuit output depends not only on the populations produced by individual pulses but also on the relative phase of the intermediate superposition. The weak performance of B5 variable rotations therefore has a direct physical interpretation: population compensation without phase stabilization is insufficient when coherent pathways later recombine.

H5s/X5, BB1, and SK1 protect the complete coherent operation more effectively, with H5s/X5 providing the broadest response in the present comparison. The increase from approximately \(8\%\) to \(50\%\) in the tolerated systematic rotation-angle error shows that the broad error response of the composite gates is retained in a multigate workload. 
The three-level R branch complements this result by distinguishing active composite compensation from conditional error cancellation generated by the circuit structure; the latter is exact only when the relevant composite propagators satisfy the inverse relation.

The benchmark deliberately addresses one coherent error channel: a common fractional error in the programmed single-qubit rotation angles. The present composite gates do not correct errors in the CZ or \(\mathrm{iSWAP}\) operations, independent calibration errors on different transitions, phase offsets, detuning, leakage, or time-dependent stochastic noise. The experiment should therefore be viewed as a controlled circuit-level test of coherent-error compensation rather than as a general error-mitigation method for the processor.

The processor is not used to identify the handedness of an unknown molecule, to provide a computational speedup, or to perform a first-principles simulation of arbitrary molecular dynamics. Its role is to realize prescribed chiral-interference state-transfer maps and deliberately expose them to a selected control error. This restricted setting is nevertheless useful because it tests composite gates in a physically meaningful phase-sensitive protocol rather than in isolation.

Natural extensions include independent amplitude errors on the different rotations, detuning and phase offsets, leakage, and stochastic noise. Composite or dynamically corrected constructions could also be developed for the entangling operations, which are the main limitation apart from SPAM. Larger rotational manifolds, unequal transition strengths, and magnetic-sublevel degeneracies would provide more demanding chiral-interference workloads, while native qutrit or qudit hardware could represent cyclic molecular couplings more directly.

\section{Conclusion\label{sec:conclusion}}

We have constructed and experimentally implemented compact gate-native circuits for three- and four-level interferometric chiral resolution. 
The circuits reproduce the prescribed enantiomer-dependent output populations with probabilities of up to 98\%, without error mitigation of SPAM and other errors.
We then used these circuits as phase-sensitive benchmarks for several composite-gate constructions under a systematic rotation-angle error. 
Constant-rotation composite gates preserve the required interference over a substantially broader error range than elementary rotations, whereas variable-rotation sequences that do not stabilize the relevant phases provide weaker protection. 
The three-level circuit additionally separates active composite-gate compensation from conditional robustness generated by circuit symmetry and makes explicit the exact-inverse requirement for preserving that cancellation. 
The results demonstrate composite-gate protection at the level of a complete physical interference protocol rather than only at the level of an isolated gate.

\section{Acknowledgments}

This research is supported by the Bulgarian national plan for recovery and resilience, contract BG-RRP-2.004-0008-C01 (SUMMIT: Sofia University Marking Momentum for Innovation and Technological Transfer), project number 3.1.4, and by the QuantERA project QUARCKS.

\appendix

\section{Single-gate propagator expansions\label{sec:a1}}
\label{app:single-gate-errors}

For a resonant pulse, let
$R_{\phi}(\theta)=\exp[-i\theta\sigma_{\phi}/2]$, with
$\sigma_{\phi}=\cos\phi\,\sigma_x+\sin\phi\,\sigma_y$, and model the
common rotation-angle error by $\theta\mapsto(1+\varepsilon)\theta$.
Then
\be
\begin{aligned}
R_{\phi}[(1+\varepsilon)\theta]
={}&
R_{\phi}(\theta)
\biggl[
\mathbf{I}
-\frac{i\theta}{2}\varepsilon\sigma_{\phi}
-\frac{\theta^2}{8}\varepsilon^2\mathbf{I}\\
&\hspace{1.3cm}
+\frac{i\theta^3}{48}\varepsilon^3\sigma_{\phi}
+\mathcal{O}(\varepsilon^4)
\biggr].
\end{aligned}
\label{eq:app-single-pulse-expansion}
\ee
For a chronological sequence
$(\theta_1)_{\phi_1}\cdots(\theta_N)_{\phi_N}$,
$U_{\rm CP}(\varepsilon)=R_{\phi_N}[(1+\varepsilon)\theta_N]\cdots
R_{\phi_1}[(1+\varepsilon)\theta_1]$.  With
$U_0=U_{\rm CP}(0)$, the target-relative propagator
$\mathcal{E}=U_0^\dagger U_{\rm CP}$ is expanded as
$\mathcal{E}=\mathbf{I}+\varepsilon E_1+\varepsilon^2E_2+\cdots$.
Defining
$V_k=R_{\phi_k}(\theta_k)\cdots R_{\phi_1}(\theta_1)$,
$V_0=\mathbf{I}$, and
$\widetilde{\sigma}_k=V_{k-1}^\dagger\sigma_{\phi_k}V_{k-1}$ gives
\be
\begin{aligned}
\mathcal{E}(\varepsilon)
={}&\mathbf{I}-\frac{i\varepsilon}{2}\sum_k\theta_k\widetilde{\sigma}_k-\frac{\varepsilon^2}{8}\sum_k\theta_k^2\mathbf{I}\\
&-\frac{\varepsilon^2}{4}\sum_{j>k}\theta_j\theta_k\widetilde{\sigma}_j\widetilde{\sigma}_k+\mathcal{O}(\varepsilon^3).
\end{aligned}
\label{eq:app-general-error-expansion}
\ee
A full-rotation composite gate compensating through order $n$ obeys
$E_1=\cdots=E_n=0$, whereas a variable-rotation sequence need only
cancel derivatives of a selected transition probability.

\subsection{Elementary rotations}

For an elementary $x$ rotation,
$\mathcal{E}_{\rm el}=\exp(-i\theta\varepsilon\sigma_x/2)$, so
\begin{subequations}
\label{eq:app-elementary-errors}
\begin{align}
\mathcal{E}_{{\rm el},\pi/2}&=\mathbf{I}-\frac{i\pi}{4}\varepsilon\sigma_x-\frac{\pi^2}{32}\varepsilon^2\mathbf{I}+\frac{i\pi^3}{384}\varepsilon^3\sigma_x+\mathcal{O}(\varepsilon^4),
\label{eq:app-elementary-halfpi}\\
\mathcal{E}_{{\rm el},\pi}&=\mathbf{I}-\frac{i\pi}{2}\varepsilon\sigma_x-\frac{\pi^2}{8}\varepsilon^2\mathbf{I}+\frac{i\pi^3}{48}\varepsilon^3\sigma_x+\mathcal{O}(\varepsilon^4).
\label{eq:app-elementary-pi}
\end{align}
\end{subequations}

\subsection{Variable-rotation B5 sequences}

For the variable-rotation sequence
\be
\mathrm{B5}_{\pi/2}=(\pi/2)_0(\pi)_{3\pi/8}(\pi)_{3\pi/2} (\pi)_{11\pi/8}(\pi/2)_0 ,
\label{eq:app-B5-halfpi-sequence}
\ee
the target-relative propagator is
\be
\begin{aligned}
\mathcal{E}_{{\rm B5},\pi/2}={}&\mathbf{I}+i\varepsilon(a_x\sigma_x+a_z\sigma_z)\\
&+\varepsilon^2\left[-\frac{\pi^2}{16}(7-4\sqrt2)\mathbf{I}+i\frac{\pi^2}{16}\sigma_y\right]+\mathcal{O}(\varepsilon^3),
\end{aligned}
\label{eq:app-B5-halfpi-error}
\ee
where the coefficients $a_x$ and $a_z$ are
\begin{subequations}
\label{eq:app-B5-halfpi-coefficients}
\begin{align}
a_x&=\frac{\pi}{4}[\sqrt{4-2\sqrt2}+\sqrt2-1]\approx 1.175,\label{eq:app-B5-ax}\\
a_z&=\frac{\pi}{4}[\sqrt{4-2\sqrt2}+1-\sqrt2]\approx 0.525 .
\label{eq:app-B5-az}
\end{align}
\end{subequations}
Nevertheless,
\be
\begin{aligned}
P_{{\rm B5},\pi/2}&=\frac12-\frac12\sin^8(\pi\varepsilon/2)\\
&=\frac12-\frac{\pi^8}{512}\varepsilon^8+\mathcal{O}(\varepsilon^{10}).
\end{aligned}
\label{eq:app-B5-halfpi-population}
\ee

For
$\mathrm{B5}_{\pi}=(\pi)_0(\pi)_{4\pi/5}(\pi)_{2\pi/5}(\pi)_{4\pi/5}(\pi)_0$,
\begin{subequations}
\label{eq:app-B5-pi-error}
\begin{align}
\mathcal{E}_{{\rm B5},\pi}&=\mathbf{I}-i\kappa_\pi\varepsilon^2\sigma_z+\mathcal{O}(\varepsilon^4),
\label{eq:app-B5-pi-propagator}\\
\kappa_\pi&=\frac{\pi^2}{16}\left[\sqrt{10+2\sqrt5}+\sqrt{10-2\sqrt5}\right]\approx 3.797 .
\label{eq:app-B5-kappa}
\end{align}
\end{subequations}
Its isolated population is
\be
\begin{aligned}
P_{{\rm B5},\pi}&=1-\sin^{10}(\pi\varepsilon/2)\\
&=1-\frac{\pi^{10}}{1024}\varepsilon^{10}+\mathcal{O}(\varepsilon^{12}).
\end{aligned}
\label{eq:app-B5-pi-population}
\ee

\subsection{SK1 sequences}

The SK1 sequence is
$\mathrm{SK1}(\theta)=(\theta)_0(2\pi)_\phi(2\pi)_{-\phi}$, with $\phi=\arccos[-\theta/(4\pi)]$.  
The first-order error cancels and
\be
\mathcal{E}_{\rm SK1}=\mathbf{I}-i\pi^2\sin(2\phi)\bigl(\sin\theta\,\sigma_y+\cos\theta\,\sigma_z\bigr)\varepsilon^2+\mathcal{O}(\varepsilon^3).
\label{eq:app-SK1-general}
\ee
For $\theta=\pi/2$, $\cos\phi=-1/8$, and expansion through third order
gives the propagator and isolated transition probability
\begin{subequations}
\label{eq:app-SK1-halfpi}
\begin{align}
\mathcal{E}_{{\rm SK1},\pi/2}
={}&\mathbf{I}+i\frac{3\sqrt7\pi^2}{32}\varepsilon^2\sigma_y \notag\\
&-i\frac{\pi^3}{128}\left(21\sigma_x+3\sqrt7\,\sigma_z\right)\varepsilon^3+\mathcal{O}(\varepsilon^4),
\label{eq:app-SK1-halfpi-propagator}\\
P_{{\rm SK1},\pi/2}={}&\frac12+\frac{21\pi^3}{128}\varepsilon^3-\frac{21\pi^5}{1024}\varepsilon^5-\frac{989\pi^7}{81920}\varepsilon^7+\mathcal{O}(\varepsilon^9).
\label{eq:app-SK1-halfpi-population}
\end{align}
\end{subequations}
Thus the complete propagator error begins at order $\varepsilon^2$,
whereas the deviation from the nominal half-population value begins at
order $\varepsilon^3$.

For $\theta=\pi$, $\cos\phi=-1/4$, and
\begin{subequations}
\label{eq:app-SK1-pi}
\begin{align}
\mathcal{E}_{{\rm SK1},\pi}={}&\mathbf{I}-i\frac{\sqrt{15}\pi^2}{8}\varepsilon^2\sigma_z \notag\\
&-i\frac{\pi^3}{16}\left(5\sigma_x+\sqrt{15}\,\sigma_y\right)\varepsilon^3+\mathcal{O}(\varepsilon^4), 
\label{eq:app-SK1-pi-propagator}\\
P_{{\rm SK1},\pi}={}&1-\frac{5\pi^6}{32}\varepsilon^6+\frac{25\pi^8}{256}\varepsilon^8+\mathcal{O}(\varepsilon^{10}).
\label{eq:app-SK1-pi-population}
\end{align}
\end{subequations}
The second-order term in the propagator is longitudinal and therefore
changes only the phase of the ideal transferred state; the isolated
transition-probability error begins at sixth order.

\subsection{BB1 sequences}

The BB1 sequence is
$\mathrm{BB1}(\theta)=(\theta)_0(\pi)_\phi(2\pi)_{3\phi}(\pi)_\phi$,
with $\phi=\arccos[-\theta/(4\pi)]$.  
Here $E_1=E_2=0$, and direct expansion gives
\begin{subequations}
\label{eq:app-BB1-errors}
\begin{align}
\mathcal{E}_{{\rm BB1},\pi/2}={}&\mathbf{I}-\frac{i\pi^3}{512}(21\sigma_x+3\sqrt7\,\sigma_z)\varepsilon^3+\mathcal{O}(\varepsilon^4)\nonumber\\
\simeq{}&\mathbf{I}-i(1.272\,\sigma_x+0.481\,\sigma_z)\varepsilon^3,\label{eq:app-BB1-halfpi}\\
\mathcal{E}_{{\rm BB1},\pi}={}&\mathbf{I}-\frac{i\pi^3}{64}(5\sigma_x+\sqrt{15}\,\sigma_y)\varepsilon^3+\mathcal{O}(\varepsilon^4)\nonumber\\
\simeq{}&\mathbf{I}-i(2.422\,\sigma_x+1.876\,\sigma_y)\varepsilon^3.
\label{eq:app-BB1-pi}
\end{align}
\end{subequations}

\subsection{X5 and H5s sequences}

The symmetric X5 sequence and its phases are
\begin{subequations}
\label{eq:app-X5-definition}
\begin{align}
\mathrm{X5}&=(\pi)_{\phi_1}(\pi)_{\phi_2}(\pi)_{\phi_3}(\pi)_{\phi_2}(\pi)_{\phi_1},
\label{eq:app-X5-sequence}\\
\phi_1&=\arcsin(1-\sqrt{5/8}) \approx 0.067161\pi,
\label{eq:app-X5-phi1}\\
\phi_2&=\arcsin[(3\sqrt{10}-2)/8] \approx 0.385370 \pi,
\label{eq:app-X5-phi2}\\
\phi_3&=2\phi_2-2\phi_1+\pi/2 \approx 1.136418 \pi.
\label{eq:app-X5-phi3}
\end{align}
\end{subequations}
The propagator expansion is
\be
\begin{aligned}
\mathcal{E}_{\rm X5}&=\mathbf{I}+i\frac{\sqrt{10}\pi^3}{32}\varepsilon^3\sigma_y+\mathcal{O}(\varepsilon^4)\\
&\simeq\mathbf{I}+i\,3.064\,\varepsilon^3\sigma_y .
\end{aligned}
\label{eq:app-X5-error}
\ee

For
$\mathrm{H5s}=(\alpha)_{\phi_1}(\pi)_{\phi_2}(\pi)_{\phi_3}
(\pi)_{\phi_2}(\alpha)_{\phi_1}$, the parameters are
\begin{subequations}
\label{eq:app-H5s-parameters}
\begin{align}
\alpha/\pi&=0.450016,
\label{eq:app-H5s-alpha}\\
\phi_1/\pi&=1.949388,
\label{eq:app-H5s-phi1}\\
\phi_2/\pi&=0.510617,
\label{eq:app-H5s-phi2}\\
\phi_3/\pi&=1.317856.
\label{eq:app-H5s-phi3}
\end{align}
\end{subequations}
They give $E_1=E_2=0$ and
\be
\begin{aligned}
\mathcal{E}_{\rm H5s}\simeq{}&\mathbf{I}+i\varepsilon^3\bigl[0.896(\sigma_x+\sigma_z)-0.287\,\sigma_y\bigr]+\mathcal{O}(\varepsilon^4).
\end{aligned}
\label{eq:app-H5s-error}
\ee
A common shift $\phi_k\mapsto\phi_k+\phi_0$ rotates the nominal axis
and transverse Pauli components but leaves the compensation order
unchanged.

\begin{table}[t]
\caption{\label{tab:app-gate-orders}
Leading rotation-error orders of the complete propagator and of the
isolated transition probability from $\ket{0}$.}
\begin{ruledtabular}
\begin{tabular}{lcc}
Gate & Propagator & Transition probability\\
\hline
Elementary & $\mathcal{O}(\varepsilon)$
& $\mathcal{O}(\varepsilon)$ or $\mathcal{O}(\varepsilon^2)$\\
B5, $\pi/2$ & $\mathcal{O}(\varepsilon)$ & $\mathcal{O}(\varepsilon^8)$\\
B5, $\pi$ & $\mathcal{O}(\varepsilon^2)$ & $\mathcal{O}(\varepsilon^{10})$\\
SK1, $\pi/2$ & $\mathcal{O}(\varepsilon^2)$ & $\mathcal{O}(\varepsilon^3)$\\
SK1, $\pi$ & $\mathcal{O}(\varepsilon^2)$ & $\mathcal{O}(\varepsilon^6)$\\
BB1 & $\mathcal{O}(\varepsilon^3)$ & $\mathcal{O}(\varepsilon^6)$\\
H5s & $\mathcal{O}(\varepsilon^3)$ & $\mathcal{O}(\varepsilon^6)$\\
X5 & $\mathcal{O}(\varepsilon^3)$ & $\mathcal{O}(\varepsilon^6)$\\
\end{tabular}
\end{ruledtabular}
\end{table}

\section{Circuit-level error expansions\label{sec:a2}}
\label{app:circuit-errors}

We now propagate the single-gate errors of Appendix~\ref{app:single-gate-errors}
through the complete three- and four-level circuits.  The CZ and \(\mathrm{iSWAP}\) gates
are assumed to be perfect, and the only imperfection is the common fractional
rotation-angle error $\varepsilon$ in the single-qubit gates.

\subsection{General circuit expansion}

Let $A=R_x(\pi/2)$ and $B=R_x(\pi)$ denote the ideal rotations in the
four-level circuit, and write their erroneous implementations as
$\widetilde A=A\mathcal E_A$ and $\widetilde B=B\mathcal E_B$, where
\begin{subequations}
\label{eq:app-single-gate-general}
\begin{align}
\mathcal E_A&=\mathbf{I}+i\varepsilon^p(x\sigma_x+y\sigma_y+z\sigma_z)+\mathcal O(\varepsilon^{p+1}),\label{eq:app-EA-general}\\
\mathcal E_B&=\mathbf{I}+i\varepsilon^q(u\sigma_x+v\sigma_y+w\sigma_z)+\mathcal O(\varepsilon^{q+1}).
\label{eq:app-EB-general}
\end{align}
\end{subequations}
For an ideal circuit $U_0$ and its erroneous implementation $\widetilde U$,
we define $\mathcal E_{\rm circ}=U_0^\dagger\widetilde U$.  If
$U_0\ket{00}=\ket{\psi_{\rm t}}$, the target population is
$P_{\rm t}=|\bra{00}\mathcal E_{\rm circ}\ket{00}|^2$.

\subsubsection{Four-level S circuit}

The ideal and erroneous propagators are
\begin{subequations}
\label{eq:app-U4S-pair}
\begin{align}
U_{4S}&=(A\otimes\mathbf{I})(\mathbf{I}\otimes B)(A\otimes\mathbf{I}),
\label{eq:app-U4S}\\
\widetilde U_{4S}&=(\widetilde A\otimes\mathbf{I}) (\mathbf{I}\otimes\widetilde B) (\widetilde A\otimes\mathbf{I}).
\label{eq:app-U4S-erroneous}
\end{align}
\end{subequations}
Because the gates acting on different qubits commute, the exact relative
error factorizes as
\be
\mathcal E_{4S}=(A^\dagger\mathcal E_AA\mathcal E_A)\otimes\mathcal E_B.
\label{eq:app-E4S-exact}
\ee
Using $A^\dagger\sigma_xA=\sigma_x$,
$A^\dagger\sigma_yA=-\sigma_z$, and
$A^\dagger\sigma_zA=\sigma_y$, its leading expansion is
\begin{align}
\mathcal E_{4S}
={}&\mathbf{I}
+i\varepsilon^p
\left[
2x\sigma_x^{(1)}+(y+z)\sigma_y^{(1)}+(z-y)\sigma_z^{(1)}\right]\nonumber\\
&+i\varepsilon^q\left[ u\sigma_x^{(2)}+v\sigma_y^{(2)}+w\sigma_z^{(2)}\right]+\cdots .
\label{eq:app-E4S-leading}
\end{align}
The transverse errors on the two qubits populate orthogonal leakage states,
so
\be
1-P_{4S}=\left[4x^2+(y+z)^2\right]\varepsilon^{2p}+(u^2+v^2)\varepsilon^{2q}+\cdots .
\label{eq:app-P4S-general}
\ee

\subsubsection{Four-level R circuit}

With $C={\rm CZ}$,
\begin{subequations}
\label{eq:app-U4R-pair}
\begin{align}
U_{4R}&=(A\otimes\mathbf{I})C(\mathbf{I}\otimes B)(A\otimes\mathbf{I}),
\label{eq:app-U4R}\\
\widetilde U_{4R}&=(\widetilde A\otimes\mathbf{I})C(\mathbf{I}\otimes\widetilde B)(\widetilde A\otimes\mathbf{I}).
\label{eq:app-U4R-erroneous}
\end{align}
\end{subequations}
The error of the final $\pi/2$ gate is transformed through the preceding
ideal circuit according to
$\sigma_x^{(1)}\mapsto-\sigma_x^{(1)}\sigma_z^{(2)}$,
$\sigma_y^{(1)}\mapsto\sigma_z^{(1)}\sigma_z^{(2)}$, and
$\sigma_z^{(1)}\mapsto\sigma_y^{(1)}$.  Therefore
\begin{align}
\mathcal E_{4R}={}&\mathbf{I}+i\varepsilon^p\Bigl\{ x[\sigma_x^{(1)}-\sigma_x^{(1)}\sigma_z^{(2)}]\nonumber\\
&\quad+y[\sigma_y^{(1)}+\sigma_z^{(1)}\sigma_z^{(2)}]+z[\sigma_z^{(1)}+\sigma_y^{(1)}]\Bigr\}\nonumber\\
&+i\varepsilon^q[u\sigma_x^{(2)}+v\sigma_y^{(2)}+w\sigma_z^{(2)}]+\cdots .
\label{eq:app-E4R-leading}
\end{align}
The two terms proportional to $x$ cancel on $\ket{00}$, and hence
\be
1-P_{4R}=(y+z)^2\varepsilon^{2p}+(u^2+v^2)\varepsilon^{2q}+\cdots .
\label{eq:app-P4R-general}
\ee

\subsubsection{Three-level circuits}

For the three-level circuit it is important to retain the phase carried by the implemented exchange. Let \(W_i=\mathrm{iSWAP}\) and define the single-qubit phase gate
\be
D=\ket{0}\!\bra{0}+i\ket{1}\!\bra{1}.
\label{eq:app-phase-gate-D}
\ee
Although \(W_i\) is not equal to SWAP as a two-qubit unitary, for the prescribed input \(\ket{00}\) and any one-qubit operation \(U\) on the first qubit one has
\be
W_i(U\otimes\mathbf I)\ket{00}=\left[\mathbf I\otimes(DUD^\dagger)\right]\ket{00}.
\label{eq:app-iswap-transport}
\ee
Equation~\eqref{eq:app-iswap-transport} is a state-specific transport identity, not a unitary equivalence between \(\mathrm{iSWAP}\) and SWAP. In particular,
\(D R_y(\theta)D^\dagger=R_x(-\theta)\), so the initial \(R_y(\pi/2)\) operation is transported by \(\mathrm{iSWAP}\) to an effective \(R_x(-\pi/2)\) operation on the second qubit. After the corresponding fixed phase conjugation of the composite-gate error vector, the leading target-population error of the S branch is
\be
1-P_{3S}=\left[4x^2+(y+z)^2\right]\varepsilon^{2p}+\cdots .
\label{eq:app-P3S-general}
\ee
For the R branch, let \(\widetilde A_y\) denote the erroneous first composite operation and \(\widetilde A_{x,+}\) the erroneous final operation. Exact cancellation occurs if and only if the final propagator is the inverse of the transported first propagator,
\be
\widetilde A_{x,+}=\left(D\widetilde A_yD^\dagger\right)^\dagger .
\label{eq:app-exact-inverse-condition}
\ee
Under this condition,
\begin{align}
\widetilde U_{3R}\ket{00}&=\left[\mathbf I\otimes\widetilde A_{x,+}(D\widetilde A_yD^\dagger)\right]\ket{00}\nonumber\\
&=\ket{00},
\label{eq:app-U3R-exact}
\end{align}
and therefore \(P_{3R}=1\) to all orders in the common error. Importantly, programming a composite sequence with the opposite nominal target angle does not automatically satisfy Eq.~\eqref{eq:app-exact-inverse-condition}. For a product of pulses, the exact inverse also reverses the pulse order. Symmetric composite constructions such as H5s can satisfy this relation directly, whereas asymmetric constructions such as B5 generally do not. The actual \(P_{3R}\) error orders for the implementations used in Fig.~\ref{fig:3ls_experiment} are given below and summarized in Table~\ref{tab:app-circuit-orders}.

\subsection{Elementary rotations}

For the elementary $\pi/2$ gate,
$x=-\pi/4$, $y=z=0$, and $p=1$.  For the elementary $\pi$ gate,
$u=-\pi/2$, $v=w=0$, and $q=1$.  Direct multiplication gives the exact
populations
\begin{subequations}
\label{eq:app-elementary-exact}
\begin{align}
P_{4S}^{\rm el}
&=\cos^4(\pi\varepsilon/2),
\label{eq:app-P4S-elementary-exact}\\
P_{4R}^{\rm el}
&=P_{3S}^{\rm el}
=\cos^2(\pi\varepsilon/2),
\label{eq:app-P4R-P3S-elementary-exact}\\
P_{3R}^{\rm el}&=1.
\label{eq:app-P3R-elementary-exact}
\end{align}
\end{subequations}
Thus
\begin{subequations}
\label{eq:app-elementary-leading}
\begin{align}
1-P_{4S}^{\rm el}
&=\frac{\pi^2}{2}\varepsilon^2
+\mathcal O(\varepsilon^4),
\label{eq:app-P4S-elementary-leading}\\
1-P_{4R}^{\rm el}
&=1-P_{3S}^{\rm el}
=\frac{\pi^2}{4}\varepsilon^2
+\mathcal O(\varepsilon^4).
\label{eq:app-P4R-P3S-elementary-leading}
\end{align}
\end{subequations}

\subsection{Variable-rotation B5 circuits}

The variable-rotation B5 sequences do not satisfy the full-propagator
conditions used in Eqs.~\eqref{eq:app-EA-general} and
\eqref{eq:app-EB-general}; their circuit errors must therefore be obtained
by direct multiplication of the pulse sequences.  For the B5 $\pi/2$ and
$\pi$ sequences listed in Table~\ref{tab:composites}, the leading complete-
circuit expansions are
\begin{subequations}
\label{eq:app-B5-circuits}
\begin{align}
1-P_{3S}^{\rm B5}
&=\pi^2\left(1-\frac{1}{\sqrt2}\right)\varepsilon^2
+\mathcal O(\varepsilon^4),
\label{eq:app-P3S-B5}\\
1-P_{4S}^{\rm B5}
&=\pi^2\left(1-\frac{1}{\sqrt2}\right)\varepsilon^2
+\mathcal O(\varepsilon^4),
\label{eq:app-P4S-B5}\\
1-P_{4R}^{\rm B5}
&=\pi^2\left(1-\frac{1}{\sqrt2}\right)\varepsilon^2
+\mathcal O(\varepsilon^4).
\label{eq:app-P4R-B5}
\end{align}
\end{subequations}
For the nominal opposite-angle B5 implementation used in the three-level R branch, the sequence is asymmetric and does not satisfy Eq.~\eqref{eq:app-exact-inverse-condition}. Direct multiplication gives
\be
1-P_{3R}^{\rm B5}=\pi^2\left(1-\frac{1}{\sqrt2}\right)\varepsilon^2 +\mathcal O(\varepsilon^4).
\label{eq:app-P3R-B5}
\ee
where $\pi^2(1-1/\sqrt2)\simeq2.891$.  The B5 $\pi$ pulse does not
contribute at this order because its isolated population error begins at
order $\varepsilon^{10}$.  The quadratic circuit error is generated by the
uncompensated phase of the variable $\pi/2$ rotation, despite its
order-$\varepsilon^8$ isolated population response.  The exact value \(P_{3R}^{\rm B5}=1\) would be recovered only if the final sequence were implemented as the exact inverse, including reversal of pulse order.

\subsection{SK1 circuits}

For the SK1 $\pi/2$ gate,
$x=z=0$, $y=3\sqrt7\pi^2/32$, and $p=2$.  The leading error of the
SK1 $\pi$ gate is longitudinal,
$u=v=0$, $w=-\sqrt{15}\pi^2/8$, and therefore does not contribute to
the leading target-population error.  Equations~\eqref{eq:app-P4S-general}--
\eqref{eq:app-P3S-general} give
\begin{subequations}
\label{eq:app-SK1-circuits}
\begin{align}
1-P_{3S}^{\rm SK1}
&=\frac{63\pi^4}{1024}\varepsilon^4+\cdots,
\label{eq:app-P3S-SK1}\\
1-P_{4S}^{\rm SK1}
&=\frac{63\pi^4}{1024}\varepsilon^4+\cdots,
\label{eq:app-P4S-SK1}\\
1-P_{4R}^{\rm SK1}
&=\frac{63\pi^4}{1024}\varepsilon^4+\cdots.
\label{eq:app-P4R-SK1}
\end{align}
\end{subequations}
For the SK1 realization used in the three-level R branch, the nominal opposite-angle sequence is not the exact inverse of the transported first sequence. Its leading deviation is
\be
1-P_{3R}^{\rm SK1}=\frac{63\pi^4}{1024}\varepsilon^4+\cdots .
\label{eq:app-P3R-SK1}
\ee
where $63\pi^4/1024\simeq5.993$.  Exact all-order cancellation would again require the explicit inverse sequence.
Here, $P_{3R}^{\rm SK1}=1$ when the last gate is the exact inverse of the
first.

\subsection{BB1 circuits}

For the BB1 $\pi/2$ gate, define
$a_h=21\pi^3/512$ and $b_h=3\sqrt7\pi^3/512$.  Then
$x=-a_h$, $y=0$, $z=-b_h$, and $p=3$.  For the BB1 $\pi$ gate, let
$a_p=5\pi^3/64$ and $b_p=\sqrt{15}\pi^3/64$, so
$u=-a_p$, $v=-b_p$, $w=0$, and $q=3$.  
Direct expansion of the complete BB1 circuits through eighth order gives
\begin{subequations}
\label{eq:app-BB1-eighth}
\begin{align}
1-P_{3S}^{\rm BB1}
={}&
6.700\,\varepsilon^6
+0.363\,\varepsilon^7
-5.131\,\varepsilon^8
+\mathcal O(\varepsilon^9),
\label{eq:app-P3S-BB1-eighth}\\
1-P_{4S}^{\rm BB1}
={}&
16.089\,\varepsilon^6
+0.363\,\varepsilon^7
-28.296\,\varepsilon^8
+\mathcal O(\varepsilon^9),
\label{eq:app-P4S-BB1-eighth}\\
1-P_{4R}^{\rm BB1}
={}&
9.620\,\varepsilon^6
+0.363\,\varepsilon^7
-24.306\,\varepsilon^8
+\mathcal O(\varepsilon^9).
\label{eq:app-P4R-BB1-eighth}
\end{align}
\end{subequations}
For the BB1 realization used in the three-level R branch, direct expansion gives \(1-P_{3R}^{\rm BB1}=b_h^2\varepsilon^6+\cdots=(63\pi^6/262144)\varepsilon^6+\cdots\simeq0.231\,\varepsilon^6+\cdots\). Thus BB1 strongly suppresses the symmetry-breaking error, but the cancellation is not exact unless the final sequence is implemented as the explicit inverse.

\subsection{H5s/X5 circuits}

The phases listed in Appendix~\ref{app:single-gate-errors} implement the
H5s and X5 rotations about the $y$ axis.  In the four-level circuit all
phases are shifted by $-\pi/2$ to obtain the required $x$ rotations.  Under
this shift, the H5s error vector becomes
\be
\mathcal E_{\rm H5s}^{(x)}
\simeq
\mathbf{I}+i\varepsilon^3
\left(
-0.287\,\sigma_x
-0.896\,\sigma_y
+0.896\,\sigma_z
\right)
+\mathcal O(\varepsilon^4).
\label{eq:app-H5s-x-error}
\ee
Thus $x\simeq-0.287$, $y\simeq-0.896$, $z\simeq0.896$, and $p=3$.
The shifted X5 error is
$\mathcal E_{\rm X5}^{(x)}
=\mathbf{I}+i(\sqrt{10}\pi^3/32)\varepsilon^3\sigma_x
+\mathcal O(\varepsilon^4)$, so
$u=\sqrt{10}\pi^3/32$, $v=w=0$, and $q=3$.
The H5s and X5 contributions are
\begin{subequations}
\label{eq:app-H5s-X5-contributions}
\begin{align}
C_{\rm H5s}
&=4x^2+(y+z)^2\simeq0.330,
\label{eq:app-H5s-contribution}\\
C_{\rm X5}
&=u^2+v^2=\frac{5\pi^6}{512}\simeq9.389.
\label{eq:app-X5-contribution}
\end{align}
\end{subequations}

Using the full-precision H5s parameters in
Appendix~\ref{app:single-gate-errors} and the analytic X5 phases, direct
expansion of the complete circuits through eighth order gives
\begin{subequations}
\label{eq:app-H5s-X5-eighth}
\begin{align}
1-P_{3S}^{\rm H5s}
\simeq{}&
0.330\,\varepsilon^6
-4.992\,\varepsilon^7
+17.546\,\varepsilon^8
+\mathcal O(\varepsilon^9),
\label{eq:app-P3S-H5s-eighth}\\
1-P_{4S}^{\rm H5s/X5}
\simeq{}&
9.719\,\varepsilon^6
-4.992\,\varepsilon^7
-110.554\,\varepsilon^8
+\mathcal O(\varepsilon^9),
\label{eq:app-P4S-H5sX5-eighth}\\
1-P_{4R}^{\rm H5s/X5}
\simeq{}&
9.389\,\varepsilon^6
-128.100\,\varepsilon^8
+\mathcal O(\varepsilon^{10}).
\label{eq:app-P4R-H5sX5-eighth}
\end{align}
\end{subequations}
There is no seventh-order contribution in the four-level R branch.  The
symmetric H5s propagator has equal diagonal elements for every
$\varepsilon$.  If it is denoted by $A$, the H5s factor in the R-target
amplitude is therefore
$A_{00}^2-A_{01}A_{10}=\det A=1$.  Hence H5s cancels exactly from that
target amplitude and
$P_{4R}^{\rm H5s/X5}=P_{{\rm X5},\pi}$.  This also explains why only even
powers occur in Eq.~\eqref{eq:app-P4R-H5sX5-eighth}.  In the four-level S
branch, the H5s and X5 population factors multiply; since both infidelities
start at sixth order, their cross term first appears at twelfth order.
Finally, because H5s is palindromic, the opposite-angle construction satisfies the exact-inverse condition, and \(P_{3R}^{\rm H5s}=1\) to all orders in the assumed common-error model.

\subsection{Summary}

\begin{table}
\caption{Leading target-population errors of the complete circuits for perfect two-qubit gates. The \(P_{3R}\) column now corresponds to the actual nominal opposite-angle composite implementations used in Fig.~\ref{fig:3ls_experiment}; exact all-order cancellation occurs only when Eq.~\eqref{eq:app-exact-inverse-condition} is satisfied. Higher-order BB1 and H5s/X5 terms are given in Eqs.~\eqref{eq:app-P3S-BB1-eighth}--\eqref{eq:app-P4R-H5sX5-eighth}.}
\label{tab:app-circuit-orders}
\begin{tabular}{lcccc}
\hline
Construction
& $1-P_{3S}$ & $1-P_{3R}$ & $1-P_{4S}$ & $1-P_{4R}$\\
\hline
Elementary
& $2.467\varepsilon^2$
& $0$
& $4.935\varepsilon^2$
& $2.467\varepsilon^2$\\[1mm]
Variable B5
& $2.891\,\varepsilon^2$
& \(2.891\,\varepsilon^2\)
& $2.891\,\varepsilon^2$
& $2.891\,\varepsilon^2$\\[1mm]
SK1
& $5.993\,\varepsilon^4$
& \(5.993\,\varepsilon^4\)
& $5.993\,\varepsilon^4$
& $5.993\,\varepsilon^4$\\[1mm]
BB1
& $6.700\,\varepsilon^6$
& \(0.231\,\varepsilon^6\)
& $16.089\,\varepsilon^6$
& $9.620\,\varepsilon^6$\\[1mm]
H5s/X5
& $0.330\,\varepsilon^6$
& $0$
& $9.719\,\varepsilon^6$
& $9.389\,\varepsilon^6$\\
\hline
\end{tabular}
\end{table}

The table displays only the leading terms. BB1 and H5s/X5 produce sixth-order circuit infidelities in the non-symmetry-protected branches because their complete single-gate propagators are corrected through second order. The variable-rotation B5 sequences have much higher-order isolated population responses, but their residual phase errors are converted by the chiral interferometers into quadratic target-population errors. In the three-level R branch, the same phase sensitivity appears as a failure of exact symmetry cancellation whenever the nominal opposite-angle composite is not the exact inverse propagator.

\bibliographystyle{unsrtnat}
\bibliography{references}

@article{PhysRevA.104.012609,
  title = {Ultrahigh-fidelity composite rotational quantum gates},
  author = {Gevorgyan, Hayk L. and Vitanov, Nikolay V.},
  journal = {Phys. Rev. A},
  volume = {104},
  issue = {1},
  pages = {012609},
  numpages = {12},
  year = {2021},
  month = {Jul},
  publisher = {American Physical Society},
  doi = {10.1103/PhysRevA.104.012609},
  url = {https://link.aps.org/doi/10.1103/PhysRevA.104.012609}
}

@article{Patterson2013Nature,
  title = {Enantiomer-specific detection of chiral molecules via microwave spectroscopy},
  author = {Patterson, David and Schnell, Melanie and Doyle, John M.},
  journal = {Nature},
  volume = {497},
  pages = {475--477},
  year = {2013},
  publisher = {Nature Publishing Group},
  doi = {10.1038/nature12150},
  url = {https://doi.org/10.1038/nature12150}
}

@article{Eibenberger2017PRL,
  title = {Enantiomer-Specific State Transfer of Chiral Molecules},
  author = {Eibenberger, Sandra and Doyle, John and Patterson, David},
  journal = {Phys. Rev. Lett.},
  volume = {118},
  issue = {12},
  pages = {123002},
  year = {2017},
  publisher = {American Physical Society},
  doi = {10.1103/PhysRevLett.118.123002},
  url = {https://link.aps.org/doi/10.1103/PhysRevLett.118.123002}
}

@article{penicillamine,
author = {Bhushan, Ravi and Kumar, Rajender},
title = {Enantioresolution of dl-penicillamine},
journal = {Biomedical Chromatography},
volume = {24},
number = {1},
pages = {66-82},
doi = {https://doi.org/10.1002/bmc.1355},
url = {https://analyticalsciencejournals.onlinelibrary.wiley.com/doi/abs/10.1002/bmc.1355},
eprint = {https://analyticalsciencejournals.onlinelibrary.wiley.com/doi/pdf/10.1002/bmc.1355},
year = {2010}
}

@article{chiralreview,
author = {Ward, Timothy J. and Ward, Karen D.},
title = {Chiral Separations: A Review of Current Topics and Trends},
journal = {Analytical Chemistry},
volume = {84},
number = {2},
pages = {626-635},
year = {2012},
doi = {10.1021/ac202892w},
note ={PMID: 22066781},
URL = {https://doi.org/10.1021/ac202892w},
eprint = {https://doi.org/10.1021/ac202892w}
}

@article{industry,
author ={Sui, Jingchen and Wang, Na and Wang, Jingkang and Huang, Xin and Wang, Ting and Zhou, Lina and Hao, Hongxun},
title  ={Strategies for chiral separation: from racemate to enantiomer},
journal  ={Chem. Sci.},
year  ={2023},
volume  ={14},
issue  ={43},
pages  ={11955-12003},
publisher  ={The Royal Society of Chemistry},
doi  ={10.1039/D3SC01630G},
url  ={http://dx.doi.org/10.1039/D3SC01630G}
}

@article{first_optical,
title = {Selective preparation of enantiomers by laser pulses: quantum model simulation for H2POSH},
journal = {Chemical Physics Letters},
volume = {306},
number = {1},
pages = {1-8},
year = {1999},
issn = {0009-2614},
doi = {https://doi.org/10.1016/S0009-2614(99)00440-6},
url = {https://www.sciencedirect.com/science/article/pii/S0009261499004406},
author = {Y. Fujimura and L. González and K. Hoki and J. Manz and Y. Ohtsuki},
}

@article{shapiro,
  title = {Coherently Controlled Asymmetric Synthesis with Achiral Light},
  author = {Shapiro, Moshe and Frishman, Einat and Brumer, Paul},
  journal = {Phys. Rev. Lett.},
  volume = {84},
  issue = {8},
  pages = {1669--1672},
  numpages = {0},
  year = {2000},
  month = {Feb},
  publisher = {American Physical Society},
  doi = {10.1103/PhysRevLett.84.1669},
  url = {https://link.aps.org/doi/10.1103/PhysRevLett.84.1669}
}

@Article{experiment,
author={Lee, JuHyeon
and Abdiha, Elahe
and Sartakov, Boris G.
and Meijer, Gerard
and Eibenberger-Arias, Sandra},
title={Near-complete chiral selection in rotational quantum states},
journal={Nature Communications},
year={2024},
month={Aug},
day={28},
volume={15},
number={1},
pages={7441},
issn={2041-1723},
doi={10.1038/s41467-024-51360-3},
url={https://doi.org/10.1038/s41467-024-51360-3}
}

@webpage{ibm_kingston,
title={Properties of the {IBM}'s {K}ingston quantum computer},
url={https://quantum.cloud.ibm.com/computers?system=ibm_kingston}
}

@article{chiral_stirap,
      title={Chiral Discrimination on Gate-Based Quantum Computers},
      author={Muhammad Arsalan Ali Akbar and Sabre Kais},
      year={2026},
      eprint={2508.18546},
      archivePrefix={arXiv},
      primaryClass={quant-ph},
      doi={10.48550/arXiv.2508.18546},
      url={https://arxiv.org/abs/2508.18546},
}

@book{BerovaNakanishi2000,
  editor    = {Berova, Nina and Nakanishi, Koji},
  title     = {Circular Dichroism: Principles and Applications},
  publisher = {Wiley-VCH},
  doi = {10.1021/ja015338n},
  address   = {New York},
  edition   = {2},
  year      = {2000}
}

@book{Nafie2011,
  author    = {Nafie, Laurence A.},
  title     = {Vibrational Optical Activity: Principles and Applications},
  doi = {10.1002/9781119976516},
  publisher = {Wiley},
  address   = {Chichester},
  year      = {2011}
}

@book{Barron2004,
  author    = {Barron, Laurence D.},
  title     = {Molecular Light Scattering and Optical Activity},
  publisher = {Cambridge University Press},
  doi = {10.1017/CBO9780511535468},
  address   = {Cambridge},
  edition   = {2},
  year      = {2004}
}

@article{Hirota2012,
  author  = {Hirota, Eizi},
  title   = {Triple Resonance for a Three-Level System of a Chiral Molecule},
  journal = {Proc. Jpn. Acad., Ser. B},
  volume  = {88},
  number  = {3},
  pages   = {120--128},
  year    = {2012},
  doi     = {10.2183/pjab.88.120}
}

@article{DomingosPerezSchnell2018,
  author  = {Domingos, S{\'e}rgio R. and P{\'e}rez, Crist{\'o}bal and Schnell, Melanie},
  title   = {Sensing Chirality with Rotational Spectroscopy},
  journal = {Annu. Rev. Phys. Chem.},
  volume  = {69},
  pages   = {499--519},
  year    = {2018},
  doi     = {10.1146/annurev-physchem-052516-050629}
}

@article{PattersonDoyle2013PRL,
  author    = {Patterson, David and Doyle, John M.},
  title     = {Sensitive Chiral Analysis via Microwave Three-Wave Mixing},
  journal   = {Phys. Rev. Lett.},
  volume    = {111},
  issue     = {2},
  pages     = {023008},
  year      = {2013},
  publisher = {American Physical Society},
  doi       = {10.1103/PhysRevLett.111.023008}
}

@article{ShubertEtAl2014,
  author  = {Shubert, V. Alvin and Schmitz, David and Patterson, David and Doyle, John M. and Schnell, Melanie},
  title   = {Identifying Enantiomers in Mixtures of Chiral Molecules with Broadband Microwave Spectroscopy},
  journal = {Angew. Chem. Int. Ed.},
  volume  = {53},
  number  = {4},
  pages   = {1152--1155},
  year    = {2014},
  doi     = {10.1002/anie.201306271}
}

@article{ShubertEtAl2016,
  author  = {Shubert, V. Alvin and Schmitz, David and P{\'e}rez, Crist{\'o}bal and Medcraft, Chris and Krin, Anna and Domingos, S{\'e}rgio R. and Patterson, David and Schnell, Melanie},
  title   = {Chiral Analysis Using Broadband Rotational Spectroscopy},
  journal = {J. Phys. Chem. Lett.},
  volume  = {7},
  number  = {2},
  pages   = {341--350},
  year    = {2016},
  doi     = {10.1021/acs.jpclett.5b02443}
}

@article{LiBruder2008,
  author    = {Li, Yong and Bruder, C.},
  title     = {Dynamic Method to Distinguish between Left- and Right-Handed Chiral Molecules},
  journal   = {Phys. Rev. A},
  volume    = {77},
  issue     = {1},
  pages     = {015403},
  year      = {2008},
  publisher = {American Physical Society},
  doi       = {10.1103/PhysRevA.77.015403}
}

@article{YeZhangLi2018,
  author    = {Ye, Chong and Zhang, Quansheng and Li, Yong},
  title     = {Real Single-Loop Cyclic Three-Level Configuration of Chiral Molecules},
  journal   = {Phys. Rev. A},
  volume    = {98},
  issue     = {6},
  pages     = {063401},
  year      = {2018},
  publisher = {American Physical Society},
  doi       = {10.1103/PhysRevA.98.063401}
}

@article{LeibscherGiesenKoch2019,
  author  = {Leibscher, Monika and Giesen, Thomas F. and Koch, Christiane P.},
  title   = {Principles of Enantio-Selective Excitation in Three-Wave Mixing Spectroscopy of Chiral Molecules},
  journal = {J. Chem. Phys.},
  volume  = {151},
  number  = {1},
  pages   = {014302},
  year    = {2019},
  doi     = {10.1063/1.5097406}
}

@article{VitanovEtAl2017STIRAP,
  author    = {Vitanov, Nikolay V. and Rangelov, Andon A. and Shore, Bruce W. and Bergmann, Klaas},
  title     = {Stimulated Raman Adiabatic Passage in Physics, Chemistry, and Beyond},
  journal   = {Rev. Mod. Phys.},
  volume    = {89},
  issue     = {1},
  pages     = {015006},
  year      = {2017},
  publisher = {American Physical Society},
  doi       = {10.1103/RevModPhys.89.015006}
}

@article{ChenEtAl2010STA,
  author    = {Chen, Xi and Lizuain, I. and Ruschhaupt, A. and Gu{\'e}ry-Odelin, D. and Muga, J. G.},
  title     = {Shortcut to Adiabatic Passage in Two- and Three-Level Atoms},
  journal   = {Phys. Rev. Lett.},
  volume    = {105},
  issue     = {12},
  pages     = {123003},
  year      = {2010},
  publisher = {American Physical Society},
  doi       = {10.1103/PhysRevLett.105.123003}
}

@article{VitanovDrewsen2019PRL,
  title     = {Highly Efficient Detection and Separation of Chiral Molecules through Shortcuts to Adiabaticity},
  author    = {Vitanov, Nikolay V. and Drewsen, Michael},
  journal   = {Phys. Rev. Lett.},
  volume    = {122},
  issue     = {17},
  pages     = {173202},
  year      = {2019},
  publisher = {American Physical Society},
  doi       = {10.1103/PhysRevLett.122.173202}
}

@article{TorosovDrewsenVitanov2020PRA,
  title     = {Efficient and Robust Chiral Resolution by Composite Pulses},
  author    = {Torosov, Boyan T. and Drewsen, Michael and Vitanov, Nikolay V.},
  journal   = {Phys. Rev. A},
  volume    = {101},
  issue     = {6},
  pages     = {063401},
  year      = {2020},
  publisher = {American Physical Society},
  doi       = {10.1103/PhysRevA.101.063401}
}

@article{TorosovDrewsenVitanov2020PRR,
  title     = {Chiral Resolution by Composite Raman Pulses},
  author    = {Torosov, Boyan T. and Drewsen, Michael and Vitanov, Nikolay V.},
  journal   = {Phys. Rev. Research},
  volume    = {2},
  issue     = {4},
  pages     = {043235},
  year      = {2020},
  publisher = {American Physical Society},
  doi       = {10.1103/PhysRevResearch.2.043235}
}

@article{LeeEtAl2022PRL,
  author    = {Lee, JuHyeon and Bischoff, Johannes and Hernandez-Castillo, A. O. and Sartakov, Boris G. and Meijer, Gerard and Eibenberger-Arias, Sandra},
  title     = {Quantitative Study of Enantiomer-Specific State Transfer},
  journal   = {Phys. Rev. Lett.},
  volume    = {128},
  pages     = {173001},
  year      = {2022},
  doi       = {10.1103/PhysRevLett.128.173001}
}

@article{LeibscherEtAl2022CommunPhys,
  author  = {Leibscher, Monika and Pozzoli, Eugenio and P{\'e}rez, Crist{\'o}bal and Schnell, Melanie and Sigalotti, Mario and Boscain, Ugo and Koch, Christiane P.},
  title   = {Full Quantum Control of Enantiomer-Selective State Transfer in Chiral Molecules Despite Degeneracy},
  journal = {Commun. Phys.},
  volume  = {5},
  pages   = {110},
  year    = {2022},
  doi     = {10.1038/s42005-022-00883-6}
}

@article{LeeEtAl2024NJP,
  author  = {Lee, JuHyeon and Bischoff, Johannes and Hernandez-Castillo, A. O. and Abdiha, Elahe and Sartakov, Boris G. and Meijer, Gerard and Eibenberger-Arias, Sandra},
  title   = {The Influence of Microwave Pulse Conditions on Enantiomer-Specific State Transfer},
  journal = {New J. Phys.},
  volume  = {26},
  pages   = {033015},
  year    = {2024},
  doi     = {10.1088/1367-2630/ad2db4}
}

@article{XuEtAl2024APL,
  author  = {Xu, Luojia and Li, Yiwen and Xu, Jianwen and Lan, Dong and Tan, Xinsheng and Yu, Yang},
  title   = {Chiral Resolution Based on Non-Adiabatic Holonomic Quantum Control via a Transmon Qutrit},
  journal = {Appl. Phys. Lett.},
  volume  = {124},
  pages   = {092601},
  year    = {2024},
  doi     = {10.1063/5.0180152}
}

@article{StefanatosEtAl2025PRA,
  author    = {Stefanatos, Dionisis and Thanopulos, Ioannis and Paspalakis, Emmanuel},
  title     = {Fast Chiral Resolution with Optimal Control},
  journal   = {Phys. Rev. A},
  volume    = {112},
  pages     = {023116},
  year      = {2025},
  doi       = {10.1103/fkmb-b4k4}
}

@article{HongEtAl2026ChemPhysChem,
  author  = {Hong, Qian-Qian and Song, Xin-Jiao and Xu, Lei and Shu, Chuan-Cun},
  title   = {Robust Control of Enantioselective State Transfer in Chiral Molecules},
  journal = {ChemPhysChem},
  volume  = {27},
  number  = {3},
  pages   = {e202500705},
  year    = {2026},
  doi     = {10.1002/cphc.202500705}
}

@article{ZhangEtAl2026AQT,
  author  = {Zhang, Guang-Hui and Song, Xue-Ke and Ye, Liu and Wang, Dong},
  title   = {Hamiltonian Inverse Engineering for Chiral Resolution in Four-Level Systems},
  journal = {Adv. Quantum Technol.},
  volume  = {9},
  number  = {5},
  pages   = {e70308},
  year    = {2026},
  doi     = {10.1002/qute.70308}
}

@article{Cummins2003,
  author  = {Cummins, Holly K. and Llewellyn, Gavin and Jones, Jonathan A.},
  title   = {Tackling systematic errors in quantum logic gates with composite rotations},
  journal = {Physical Review A},
  volume  = {67},
  number  = {4},
  pages   = {042308},
  year    = {2003},
  doi     = {10.1103/PhysRevA.67.042308}
}

@article{Xiao2006,
  author  = {Xiao, Li and Jones, Jonathan A.},
  title   = {Robust logic gates and realistic quantum computation},
  journal = {Physical Review A},
  volume  = {73},
  number  = {3},
  pages   = {032334},
  year    = {2006},
  doi     = {10.1103/PhysRevA.73.032334}
}

@article{Mount2015,
  author  = {Mount, Emily and Kabytayev, Chingiz and Crain, Stephen and
             Harper, Robin and Baek, So-Young and Vrijsen, Geert and
             Flammia, Steven T. and Brown, Kenneth R. and Maunz, Peter and
             Kim, Jungsang},
  title   = {Error compensation of single-qubit gates in a surface-electrode
             ion trap using composite pulses},
  journal = {Physical Review A},
  volume  = {92},
  number  = {6},
  pages   = {060301},
  year    = {2015},
  doi     = {10.1103/PhysRevA.92.060301}
}

@article{BrownHarrowChuang2004,
  author  = {Brown, Kenneth R. and Harrow, Aram W. and Chuang, Isaac L.},
  title   = {Arbitrarily Accurate Composite Pulse Sequences},
  journal = {Physical Review A},
  volume  = {70},
  number  = {5},
  pages   = {052318},
  year    = {2004},
  doi     = {10.1103/PhysRevA.70.052318}
}

@article{Torosov2022,
  author  = {Boyan T. Torosov and Nikolay V. Vitanov},
  title   = {Experimental Demonstration of Composite Pulses on {IBM}'s Quantum Computer},
  journal = {Physical Review Applied},
  volume  = {18},
  number  = {3},
  pages   = {034062},
  year    = {2022},
  doi     = {10.1103/PhysRevApplied.18.034062}
}

@article{torosov2011,
  title = {Smooth composite pulses for high-fidelity quantum information processing},
  author = {Torosov, Boyan T. and Vitanov, Nikolay V.},
  journal = {Phys. Rev. A},
  volume = {83},
  issue = {5},
  pages = {053420},
  numpages = {7},
  year = {2011},
  month = {May},
  publisher = {American Physical Society},
  doi = {10.1103/PhysRevA.83.053420},
  url = {https://link.aps.org/doi/10.1103/PhysRevA.83.053420}
}

@article{torosov2019,
  title = {Arbitrarily accurate variable rotations on the Bloch sphere by composite pulse sequences},
  author = {Torosov, Boyan T. and Vitanov, Nikolay V.},
  journal = {Phys. Rev. A},
  volume = {99},
  issue = {1},
  pages = {013402},
  numpages = {10},
  year = {2019},
  month = {Jan},
  publisher = {American Physical Society},
  doi = {10.1103/PhysRevA.99.013402},
  url = {https://link.aps.org/doi/10.1103/PhysRevA.99.013402}
}

@article{wimperis1994,
title = {Broadband, Narrowband, and Passband Composite Pulses for Use in Advanced NMR Experiments},
journal = {Journal of Magnetic Resonance, Series A},
volume = {109},
number = {2},
pages = {221-231},
year = {1994},
issn = {1064-1858},
doi = {https://doi.org/10.1006/jmra.1994.1159},
url = {https://www.sciencedirect.com/science/article/pii/S1064185884711594},
author = {S. Wimperis},
}

\end{document}